\documentclass[fleqn,usenatbib]{mnras}

\usepackage{newtxtext,newtxmath}

\usepackage[T1]{fontenc}

\usepackage{soul}
\usepackage{rotating}
\usepackage{pdflscape}

\DeclareRobustCommand{\VAN}[3]{#2}
\let\VANthebibliography\thebibliography
\def\thebibliography{\DeclareRobustCommand{\VAN}[3]{##3}\VANthebibliography}

\usepackage{graphicx}	
\usepackage{amsmath}
\usepackage{longtable} 

\title[Helioseismic QBO]{Spatio-temporal analysis of helioseismic quasi-biennial oscillations}

\author[Hasanzadeh et al.]
{Amir Hasanzadeh$^{1}$ \href{https://orcid.org/0000-0002-7286-1438} {\includegraphics[scale=0.4]{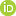}} \thanks{E-mail: amir.hasanzadeh@gmail.com}
Anne-Marie Broomhall$^{1}$ \href{https://orcid.org/0000-0002-5209-9378} {\includegraphics[scale=0.4]{orcid.png}} 
Dmitrii Kolotkov$^{1, 2}$  \href{https://orcid.org/0000-0002-0687-6172} {\includegraphics[scale=0.4]{orcid.png}}
Tishtrya Mehta$^{1}$ \href{https://orcid.org/0000-0002-4875-9142} {\includegraphics[scale=0.4]{orcid.png} }
\\
$^{1}$Centre for Fusion, Space and Astrophysics, Department of Physics, University of Warwick, Coventry CV4 7AL, UK\\
$^{2}$Engineering Research Institute \lq\lq Ventspils International Radio Astronomy Centre (VIRAC)\rq\rq, Ventspils University of Applied Sciences, Ventspils, LV-3601, Latvia
}

\date{Accepted XXX. Received YYY; in original form ZZZ}

\pubyear{2025}

\begin{document}
\label{firstpage}
\pagerange{\pageref{firstpage}--\pageref{lastpage}}
\maketitle

\begin{abstract}
Quasi-biennial oscillations (QBOs) are shorter-term periodic signals that occur alongside the dominant 11-year solar cycle.
In this study, we examine the spatial and temporal evolution of QBOs using helioseismic p-mode frequency shifts from the Global Oscillation Network Group (GONG) across solar Cycles 23 and 24 and the ascending phase of Cycle 25.
By applying wavelet analysis to frequency shifts, we studied the changes in QBO periodicities to determine whether the QBO period and amplitude vary with latitude.
Our results show that QBO periods exhibit a weak latitudinal dependence, with shorter and less persistent signals at low latitudes, while at higher latitudes the periods are nearly constant at $\sim$3 years. Cycle 24 tends to display slightly longer periods than Cycle 23, though within uncertainties.
At all latitudes, QBO amplitudes increase with mode frequency, which is consistent with previous studies. Higher amplitude QBOs are found at low latitudes, reflecting the distribution of surface magnetic activity.
The ratio of QBO to cycle amplitude is systematically higher in Cycle 24 than in Cycle 23, and above $20^\circ$ latitude the amplitude ratio is nearly uniform in Cycle 23 but shows modest variations in Cycle 24. A linear relation between QBO amplitude and cycle amplitude is found in both cycles, but with significantly different slopes, indicating that QBO amplitudes are not wholly governed by the solar cycle strength and are at least partially decoupled from it. Finally, we find no evidence that QBO period depends on QBO amplitude, consistent with a linear oscillation regime.
\end{abstract}



\section{Introduction}
The Sun's magnetic activity varies over a timescale of around 11\,years, which is known as a solar cycle \citep{2010LRSP....7....1H}. However, when plotted as a function of time, many solar activity indices show
a shorter-period variation alongside the cycle, commonly referred to as quasi-biennial oscillations (QBOs). These oscillations are typically associated with periodicities ranging from 0.6 to 4 years \citep{2005SoPh..227..155K,2007AdSpR..40.1006O,2014SSRv..186..359B, 2022MNRAS.515.2415M,2023ApJ...959...16J}. QBOs represent a significant aspect of solar activity and exhibit variations that overlay the 11-year solar cycle. 

Helioseismology uses the Sun's internal oscillations to infer conditions beneath the Sun's visible surface. The dominant oscillations used in helioseismology are acoustic oscillations, known as p modes. The p-mode frequencies have been observed to vary in phase with the solar cycle, shifting to higher values at solar maximum as the magnetic activity increases \citep[][and references therein]{1990Natur.345..779L, 1990Natur.345..322E, 1999ApJ...524.1084H, 2008AdSpR..41..861R, 2015sac..book..191B}. There is a strong correlation between the observed frequency shifts (hereafter, shifts) and solar proxies \citep {2009ApJ...695.1567J, 2015SoPh..290.3095B, 2015MNRAS.451.4360K}. Quasi-biennial signals have been detected in various helioseismic data \citep{2010ApJ...718L..19F,2011MNRAS.413.2978B, 2012A&A...539A.135S, 2013ApJ...765..100S, 2022MNRAS.515.2415M}. 

The shape of the frequency dependence of the shifts can be an important indicator of the location of the perturbation responsible for the shifts. Frequency shifts that increase monotonically with frequency are indicative of a near surface perturbation. This is because the radius of the upper turning point of the oscillations increases with frequency, and so modes with higher frequencies have more sensitivity to near surface layers \citep[e.g.][]{1990LNP...367..283G, 1990Natur.345..779L}. On the other hand, a magnetic field located deeper in the interior, such as at the base of the convection zone, would cause the amplitude of the frequency shifts to oscillate as a function of frequency \citep{1988IAUS..123..155G}. The amplitude of the QBO in the shifts appears to be less dependent on the mode frequency than the solar cycle shifts \citep {2010ApJ...718L..19F, 2022MNRAS.515.2415M}. This could indicate that the magnetic perturbation responsible for the helioseismic QBO may be spread over a larger range of depths than the solar cycle mode frequency perturbation \citep{2023ApJ...959...16J}. 

\cite {2016ApJ...828...41S} examined the latitude dependence of the p-mode frequency shifts and found that the amplitudes of the solar-cycle shifts were larger at lower latitudes, but did not consider the amplitude of the QBO.
\cite{2021ApJ...920...49I} extended this work using HMI and MDI data, identifying QBO-like periodicities in rotation rate residuals. Not only were the QBO signatures observed at depths of $0.99$, $0.95$ and $0.90R_{\odot}$ but the amplitude of the QBO was found to increase with depth, which, the authors suggest, may indicate a deeper source region for the QBO compared to the solar cycle, in agreement with \citet{2010ApJ...718L..19F, 2022MNRAS.515.2415M}. Futhermore, \citet{2021ApJ...920...49I} found that the QBO signature was present in the rotation-rate residuals at all latitudes. This is in agreement with \citet{2012ApJ...749...27V}, who found evidence for the QBO distributed over all latitudes in the NSO/Kitt Peak magnetic synoptic maps. While \citet{2020MNRAS.494.4930D} found evidence for the QBO in polar faculae. Interestingly, \citet{2020MNRAS.494.4930D} found that the amplitude of the polar faculae QBO was at a maximum at solar minimum, indicating that the QBO may persist through the solar minimum but only at higher latitudes where the more traditional activity proxies, such as sunspot number, have limited sensitivity.

The QBO has been associated with several potential mechanisms, such as two types of dynamo action, with one occurring at the bottom of the convection zone and another near the surface \citep{1998ApJ...509L..49B}, a non-linear fragment dynamo in the convection zone \citep{2000A&A...363L..13C}, and a magnetic configuration oscillation \citep{2013ApJ...765..100S}. \citet{2017NatSR...714750D} suggested that a non-linear interaction between Rossby waves and the tachocline may distort the upper boundary of the tachocline on a timescale of around two years and thus may be responsible for the QBO. Such a model is consistent with the suggestion mentioned above that a deeper source may be responsible for the QBO, demonstrating that observational constraints on the QBO may help distinguish between such models. 

We investigate how QBO periodicities and amplitudes vary with latitude by applying a wavelet analysis to helioseismic shift data from GONG. Section 2 outlines the data and methods used. In Section 3, we describe the latitudinal dependence of the 11-yr solar cycle associated frequency shifts. Section 4 presents the results of QBO period extraction across latitude bands. In Section 5, we analyse the QBO amplitude variation and its relationship to the solar cycle. Finally, in Section 6, we summarise our findings and discuss their implications for understanding solar dynamo processes.

\section{Data and Methods}
In this paper, we have used helioseismic data from GONG \citep{1996Sci...272.1284H}. Since 1995, GONG has been observing the visible solar disk from six ground-based observatories. GONG captures full-disk solar images at a rate of one image per minute \citep{1996Sci...272.1292H}. The p modes are identified by radial order ($n$), harmonic degree ($l$) and azimuthal order ($m$). 
The frequencies of the components are determined observationally up to $l$=150 using sub-series of 108 days \citep{1999ApJ...524.1084H}, obtained by combining three time series of 36 days each, which are referred to as GONG months. 
Lorentzian profiles are utilised to fit the peaks in the spectra through a minimisation scheme that relies on an initial guess to determine parameters such as the mode frequency, full width at half maximum (FWHM), and frequency error \citep{1998ESASP.418..225H}. 
The data are then processed using specific rejection criteria to eliminate anomalies and poor fits. The outcome of this process is the mrv1f data products that we used \footnote{\url{https://gong2.nso.edu/ftp/TSERIES/v1f}}. The data were filtered using the indicators within the mrv1f data files to remove frequencies obtained from bad fits.

This study used the intermediate-degree (20$\le$ $l$ $\le$150) p-mode frequencies and all available $m$ components. We have determined the change in frequency as a function of time, known as the frequency shift, by comparing the individual mode frequency (i.e. the frequency of a mode with a given $l$, $m$, $n$) observed in a specific 108\,day sub-series with a reference mode frequency. The reference mode frequency was the weighted average frequency for a given $l$, $m$, $n$ over the entire epoch. We only considered modes with frequencies below $4100\,\rm\upmu Hz$; beyond this point, errors in the parameters of the recovered p-mode, obtained by fitting a Lorentzian profile, become increasingly significant due to the increase in the width of the Lorentzian profile and the reduced signal-to-noise ratio \citep{2004A&A...413.1135S}. We rejected modes not present in all data sets, allowing us to analyse only the remaining modes referred to as "common modes" \citep{2022MNRAS.515.2415M}. These have a well-defined frequency for every GONG month within the investigated period. The first 108\,day GONG data set began on $16^\text{th}$ August 1996, while the last 108\,day GONG data set used in this analysis ended on $20^{\text{th}}$ February 2023.

Previous work has shown that the magnitude of the frequency shifts is influenced by $l$, and that this $l$-dependence is related to mode inertia \citep{1980tsp..book.....C,1990Natur.345..779L,1998MNRAS.300.1077C}. The mode inertia can be considered as a measurement of the interior mass influenced by a specific mode. 
Each mode therefore has an associated mode inertia that depends on the mode's frequency and $l$ since these determine the upper and lower bounds of the cavities in which the modes are trapped \citep{2014SSRv..186..191B}. 
We corrected the frequency shifts for this dependence using the mode inertia ratio ($Q$) which is defined as the ratio of mode mass for a given $l$ and frequency and the equivalent mode mass that an $l=0$ mode would have at the same frequency \citep{2016LRSP...13....2B}. Values of mode mass were obtained from Model S \citep{1996Sci...272.1286C}.
For each $l$, the masses are only available at specific frequency values. Therefore, we interpolated $Q$ for a given $l$ at the observed frequency for each mode. Multiplying the frequency shifts by $Q$ removed the $l$ dependence on the frequency shift magnitude \citep[e.g.][]{2001MNRAS.324..910C}. 

We calculated the corrected frequency shift time series for each mode. However, since the frequency shifts for individual modes are noisy with relatively large errors, we determined the average frequency shifts over certain ranges in mode frequency and using specific combinations of $l$ and $m$, based on their latitudinal sensitivity (as described in Section ~\ref{section_latitudes}). The average frequency shifts were obtained by determining the appropriate weighted average of the individual mode frequency shifts. Table \ref{table[nmodes]} contains the number of common modes found in each frequency and latitude range. We note there is substantial variation in the numbers and this should be kept in mind when interpreting results, particularly when considering results that cover the entire frequency or $l$ range as these will be weighted towards certain sub ranges e.g. higher $l$ and mid frequencies.

\section{Latitudinal Dependency of Frequency Shifts}\label{section_latitudes}
\begin{figure}
\begin{center}
\includegraphics[width=0.48\textwidth]{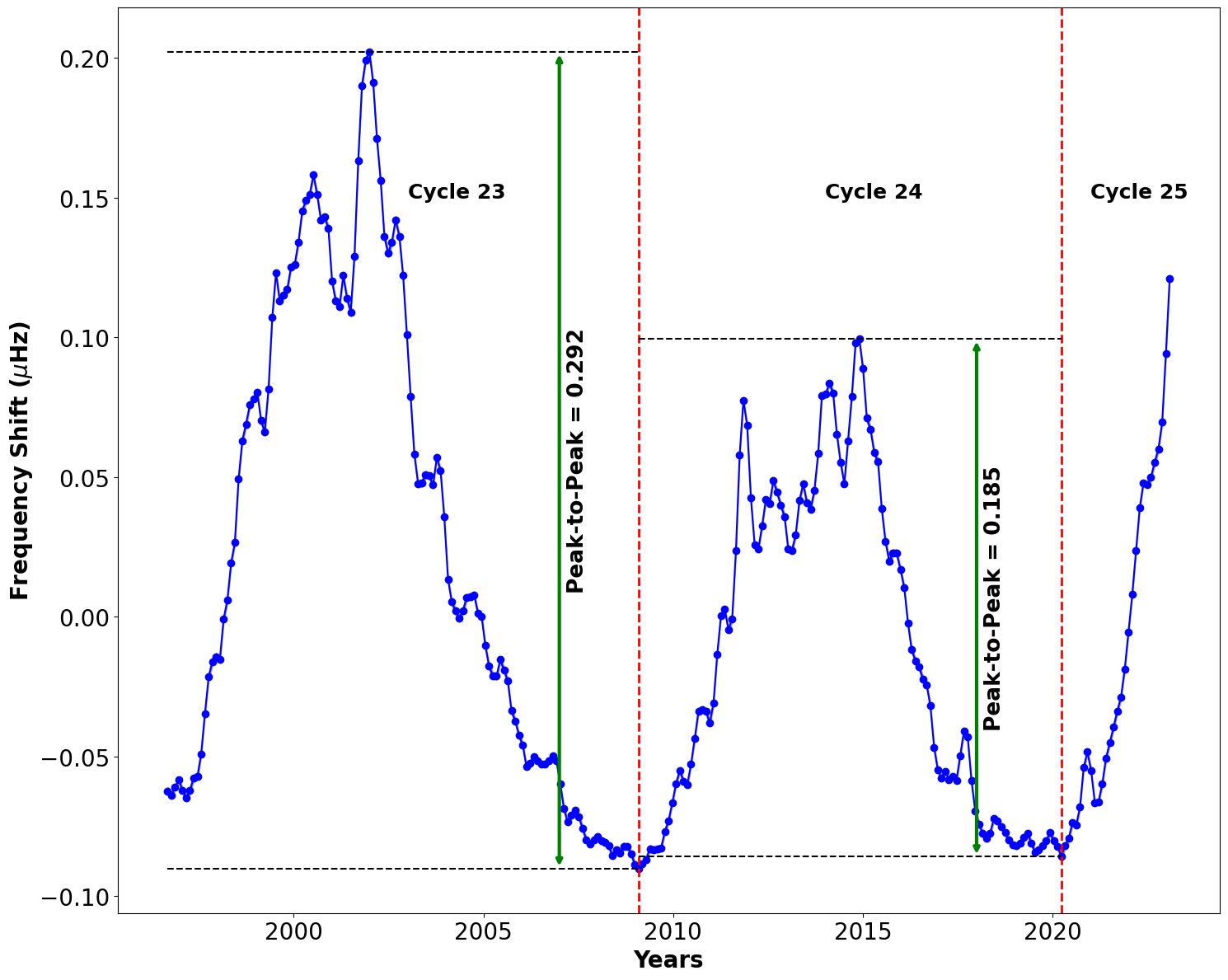}
\caption{Variation of the averaged-degree mode frequencies based on GONG data for Cycles 23 and 24 and a part of Cycle 25’s rising phase. The errors of the data are of the order of $n$Hz and are not visible in the figure. The shifts were obtained by averaging for all common modes in the frequency range 1900\,-\,4100$\,\rm\upmu Hz$ and for modes in the range $20\le l\le 150$.
The red dashed vertical lines indicate the minima of cycles. The grey horizontal lines display the minimum and maximum values, and the green solid lines show the difference between the maximum and minimum (peak-to-peak) values.}
\label{fig:shift}
\end{center}
\end{figure}

Fig. ~\ref{fig:shift} shows the frequency shifts determined by averaging the common modes over the entire range of $l$ and $m$ within a frequency range from 1900 to 4100$\,\upmu$Hz during Cycles 23 and 24, as well as the ascending branch of Cycle 25.
The shift trend aligns with the activity cycles. Thus, the negative value corresponds to the minimum and the largest positive values to the maximum of solar cycles. 
As expected, the frequency shifts increase during solar maxima and decrease at minima, in agreement with earlier helioseismic studies \citep[e.g.][]{2012MNRAS.420.1405B,2018MNRAS.480L..79H}. 

We define the peak-to-peak amplitude of the solar cycle as seen in the shifts as the difference between the maximum and minimum observed values, where the minimum values were taken from the minimum that occurs after the maximum, as demonstrated in Fig. ~\ref{fig:shift}. The errors in the individual shifts were propagated to give the error in the amplitude.

The frequency shift of a mode is highly correlated with the magnetic flux distribution across the surface \citep{2002ApJ...580.1172H} and, therefore, the latitudinal dependence of the spherical harmonic that describes each mode. For each combination of $l$ and $m$, we calculated the latitude as follows:
\begin{equation}\label{eq_theta}
\theta = \arccos\frac{m}{\sqrt{l(l+1)}}.
\end{equation}
The latitude $\theta$ represents the upper latitude to which the mode has significant sensitivity. We used this expression to categorise the previously determined shifts into different latitude bands \citep{2016ApJ...828...41S}. We averaged the shift of the modes within 10-degree bands of $\theta$. 

Near the surface, modes are reflected because of the sharp drop in density, creating an upper turning point. As the frequency of a mode increases, the upper turning point moves closer to the surface \citep{2001MNRAS.324..910C, 2012ApJ...758...43B, 2018MNRAS.480L..79H}. The frequency dependence of the location of the upper turning point means that if the perturbation is confined near the surface the magnitude of the observed shift increases with mode frequency \citep {1990Natur.345..779L,2001A&A...379..622J,2016A&A...589A.103R}. To determine the impact of mode frequency on the amplitude of the QBO, we determined the shifts by averaging over different frequency intervals.

\begin{figure}
\begin{center}
\includegraphics[width=0.48\textwidth]{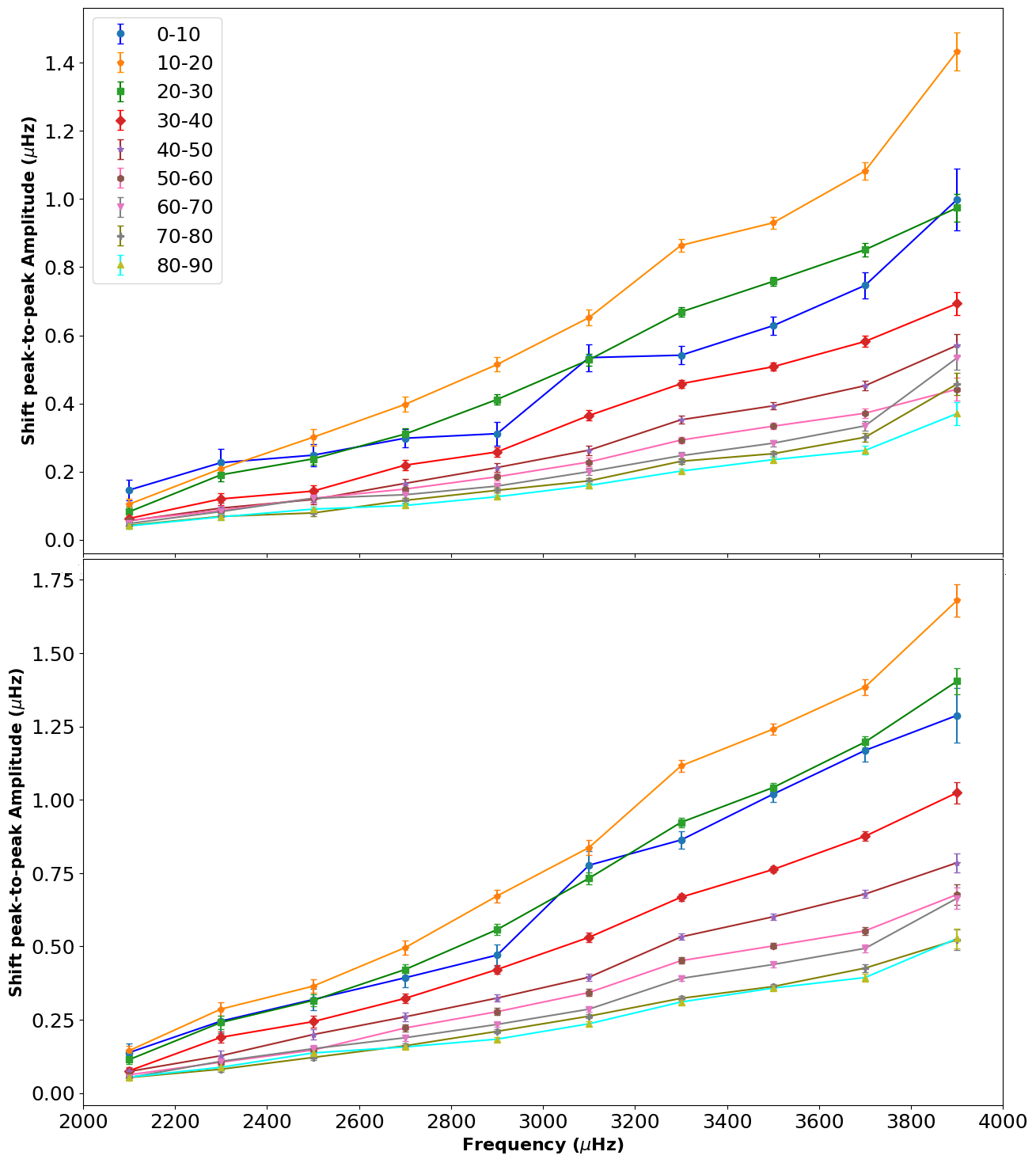}
\caption{Peak-to-peak (Max-Min) amplitude variations of shift in Cycles 23 (lower) and 24 (upper) determined by averaging over given frequency ranges and latitude bands (see legend).}
\label{fig:amp-freq}
\end{center}
\end{figure}

Using shifts like those plotted in Fig. ~\ref{fig:shift}, we calculated the peak-to-peak amplitude of the solar cycles when the shifts were determined by averaging over different frequency ranges and latitude bands for Cycles 23 and 24. These cycle amplitudes were used in Section \ref{sec:qbo_amplitudes} to determine how the amplitudes of the QBO varies from cycle to cycle in comparison to the amplitude of the solar cycle. These peak-to-peak amplitudes are shown in Fig. ~\ref{fig:amp-freq}. Although the amplitudes vary from cycle to cycle, the trend in the amplitude across latitudinal bands is similar, and, in both cases, the shift increases with rising frequency, as expected. The upward trend of the amplitude with increasing frequency confirms that higher frequency modes are more sensitive to magnetic perturbations \citep{2001A&A...379..622J}. The amplitudes are generally greater in low latitudes than in bands closer to the poles, with the $10\le \theta<20$ degree band exhibiting the highest amplitude. In contrast, the polar regions demonstrate minimal variations. This suggests a latitudinal dependence of the shifts, likely reflecting the distribution of magnetic activity. The significant mid-latitude shifts indicate that these regions are the most magnetically active during solar maxima. In contrast, the minor shifts observed near the poles suggest weaker magnetic perturbations in those areas. However, the $0\le\theta<10$ degree band deviates from this pattern as the observed amplitudes are predominantly lower than those observed in the ($10\le\theta<20$) and ($20\le\theta<30$) degree bands. This finding is consistent with \citet{2016ApJ...828...41S}.

\section{QBO periods in Latitude bands}\label{sec:qbo_periods}
\begin{figure*}
\begin{center}
\includegraphics[width=0.90\textwidth]{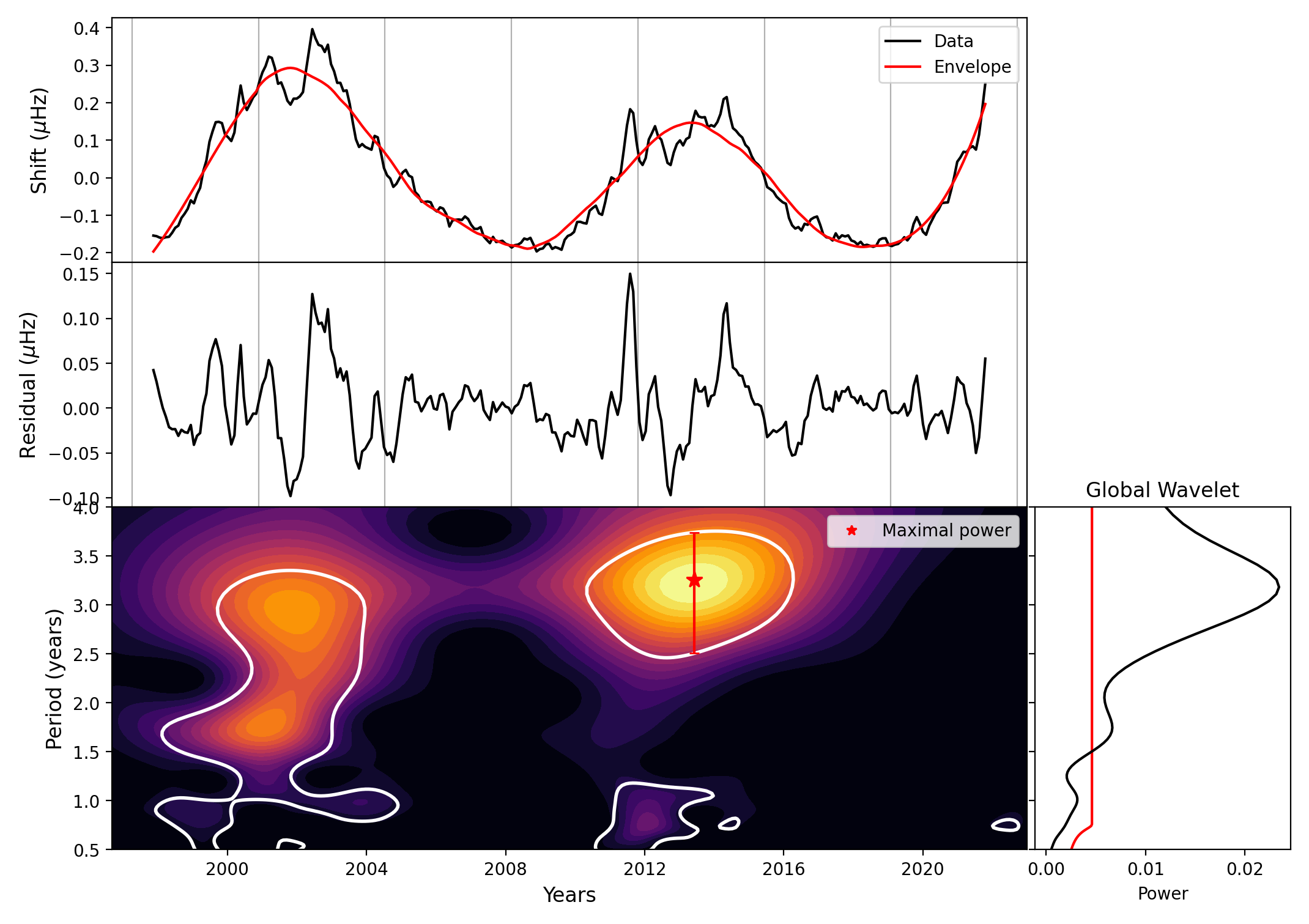}
\caption{Upper Panel: Frequency shift of the $20\le\theta < 30$ degree band for the frequency range of 2700 to $3100\,\upmu$Hz. A Savitzky-Golay filter (red line) was applied to capture the 11-year periodic variations.
Middle Panel: Shorter-term periodicity that remains after detrending the shifts by subtracting the Savitzky–Golay filtered data. Bottom left panel: Continuous wavelet transform (CWT) power spectrum of the residuals. The colours indicate normalised wavelet power and the white contours show the $95\%$ confidence levels. Bottom right panel: The corresponding global wavelet transform (GWT), which represents the time-averaged wavelet power as a function of period. The red line is the 95\% confidence level.}
\label{fig:shift-smooth}
\end{center}
\end{figure*}

The QBO is quasi-periodic, resulting in period drifts that make it unsuitable for Fourier-based analysis techniques. Therefore we retrieved the QBO periods for Cycles 23 and 24 in the latitudinal bands using the Morlet wavelet method \citep{1998BAMS...79...61T}.
Before performing this analysis, we eliminated the influence of the 11-year cycle on the shifts by subtracting a smoothed version of the shifts. This smoothed version was obtained by applying a Savitzky–Golay filter \citep{1964AnaCh..36.1627S}. A third-degree polynomial fit with a window length of 70 GONG months (about 7 years) was employed, effectively isolating the longer-term periodicities. When this smoothed version was subtracted from the shifts, only the shorter-term fluctuations (called residuals) remained (for example, see Fig. ~\ref{fig:shift-smooth}). Wavelet analyses were conducted on the residuals across the different latitudes and frequency bands. As an illustrative example, Fig. ~\ref{fig:shift-smooth} shows the  continuous wavelet transform (CWT) heatmap for the $20\le\theta < 30$ latitude band and the 2700 to 3300 $\upmu$Hz  frequency interval. To allow readers to evaluate the sharpness of the periodic peaks across all latitude bands, we provide additional heatmaps covering the full 1900–4100 $\upmu$Hz range in Appendix \ref{sec:app_figures}.

In the CWT shown in Fig. \ref{fig:shift-smooth}, the white contours in the heatmaps represent the $95\%$ confidence contours and thus highlight significant signals. Notice in Fig. \ref{fig:shift-smooth} that a significant signal was present around the maximum of cycle 23 (between 2000-2004) and the maximum of cycle 24 (2012-2016) but not at the cycle minima (around 2008 and 2020). This was a common feature of the obtained CWT plots. We therefore extracted QBO periods from the significant regions in both cycles 23 and 24. These were obtained by determining the location of the maximum amplitude within each significant region (as indicated by the star in Fig. \ref{fig:shift-smooth} for cycle 24). The width of the $95\%$ significance contours at the selected time were then utilised to calculate the periodicity error at maximum power. We note that QBO periodicities were only extracted when power above the $95\%$ significance level was observed. The periodicities obtained are given in Tables \ref{C23-period} and \ref{C24-period} for Cycles 23 and 24, respectively.

\begin{figure}
\begin{center}
\includegraphics[width=0.48\textwidth]{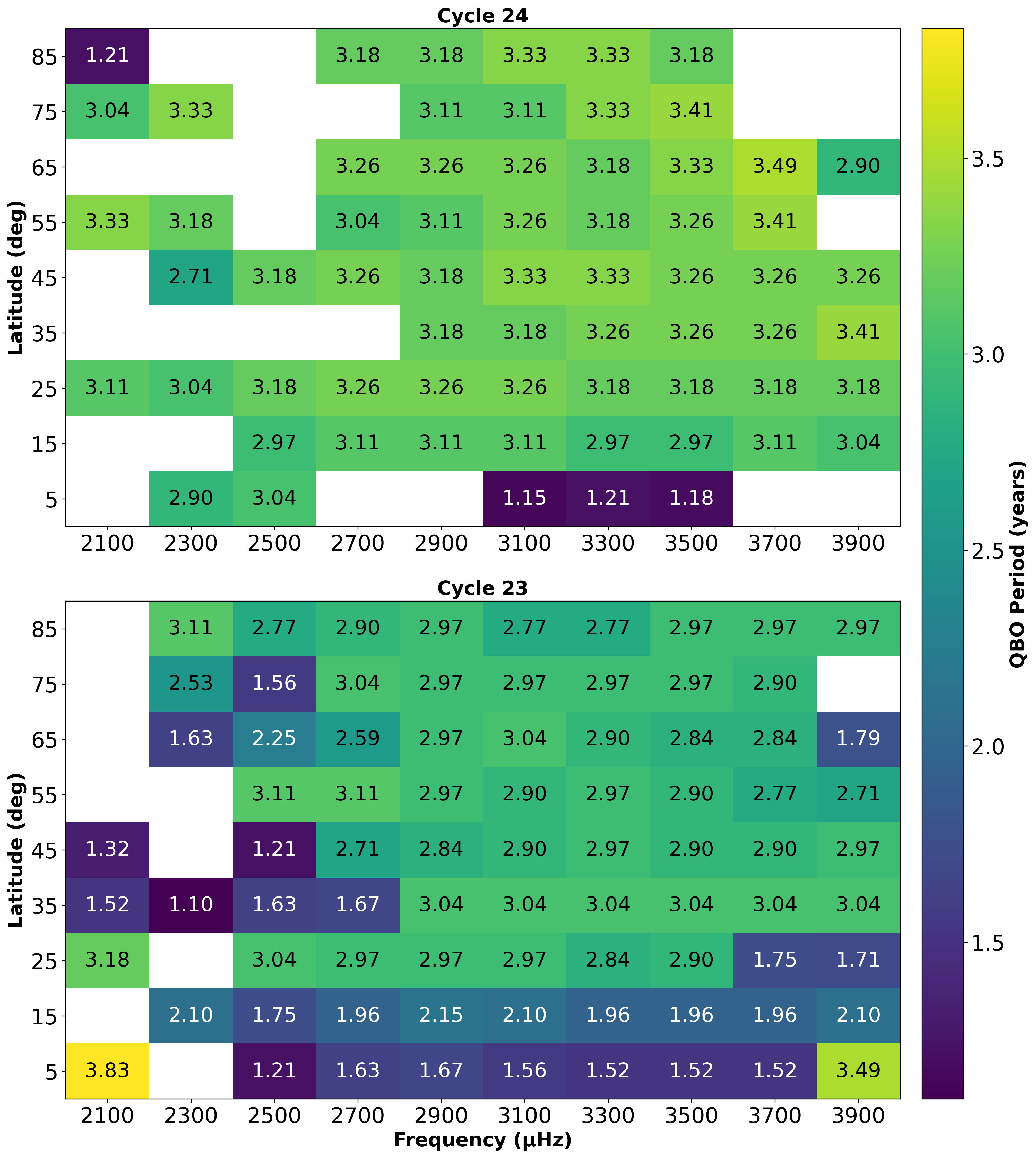}
\caption{Two-dimensional distribution of QBO periods as a function of latitude and mode frequency for Cycles 23 (bottom panel) and 24 (top panel). The colour scale is identical for both panels to allow direct comparison.}
\label{fig:QBO-period-band}
\end{center}
\end{figure}

To investigate the robustness of the detected QBO signal and its possible depth dependence, the p-mode frequency shifts were analysed in overlapping 400\,$\mu$Hz frequency bands with a 200\,$\mu$Hz overlap. This choice represents a compromise between frequency resolution and signal-to-noise ratio, and allows us to test whether the inferred QBO properties depend systematically on mode frequency.
We find that the QBO periods extracted from individual frequency bands exhibit some variability depending on latitude, cycle, and frequency range.
Figure \ref{fig:QBO-period-band} illustrates the distribution of QBO periods as a function of latitude and frequency for Cycles 23 and 24. The latitudinal periods formally span about 1 to 3.5 years, and shorter apparent periods are mainly confined to low latitudes and lower-frequency bands, particularly in Cycle 23.
The similarity of the dominant periodicity across frequencies indicates that the QBO is not intrinsically frequency dependent. This variability does not necessarily imply the presence of multiple physically distinct oscillations. Instead, it likely reflects the quasi-periodic nature of the QBO and the finite resolution of the Morlet wavelet.
We interpret this behaviour as arising from the combined effects of broader wavelet power peaks, temporal overlap of nearby periodic components, and enhanced sensitivity to noise in narrower or lower-frequency subsets.

\begin{figure}
\begin{center}
\includegraphics[width=0.48\textwidth]{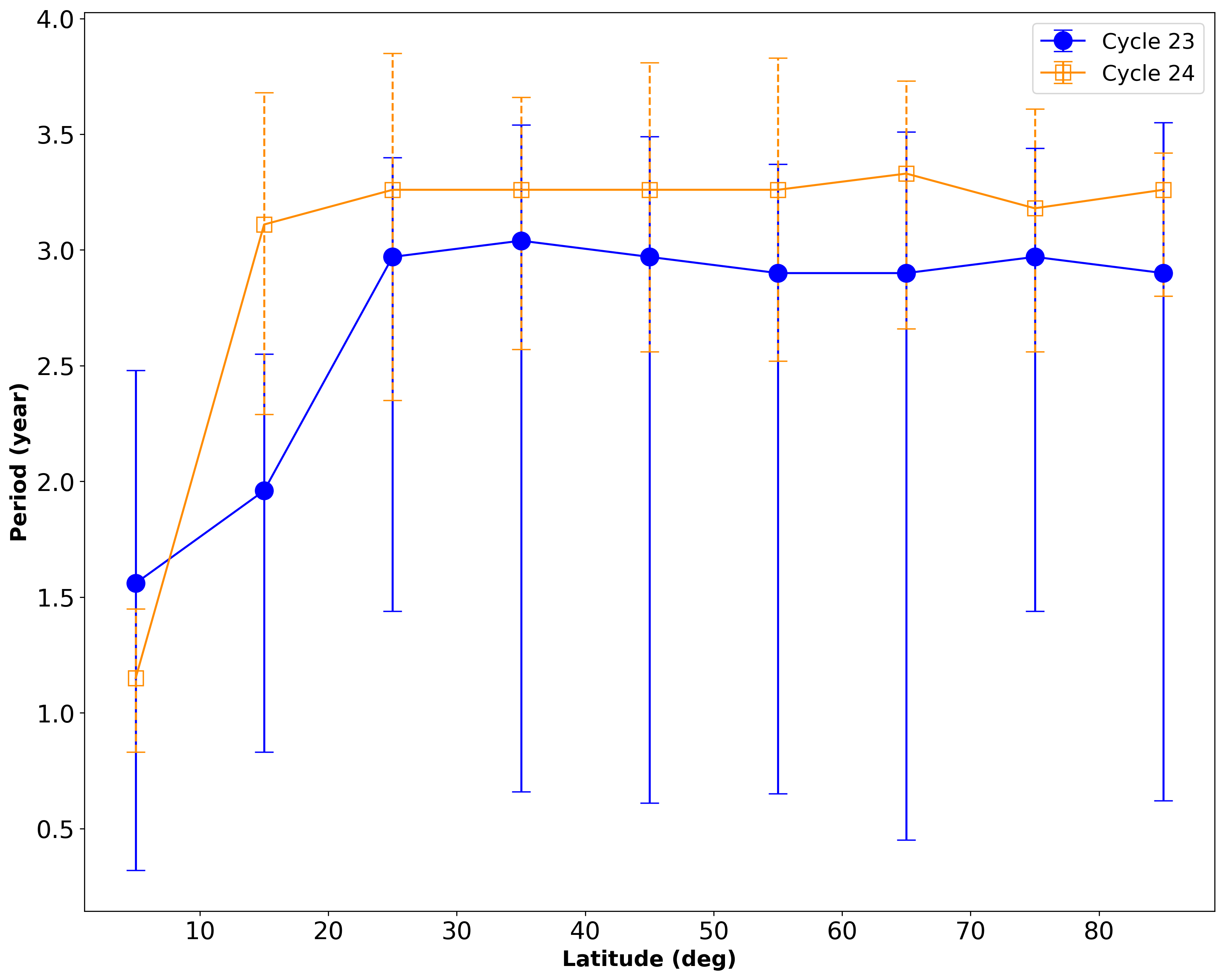}
\caption{QBO periods (in years) obtained from CWT in latitude bands for Cycles 23 (filled circles) and 24 (open squares). Frequency shifts from which the periods were obtained were determined by averaging over the entire frequency range considered.}
\label{fig:QBO-period}
\end{center}
\end{figure}

The results indicate that QBO-like signals exist across the range of frequency and latitudinal bands. In the $0\le\theta < 10$ band, although QBO periodicities were detected in a few frequency intervals, the shifts were often associated with a single-peaked maximum in activity. As a result, in some frequency ranges, no significant periodicities were detected for this latitude band (see Tables \ref{C23-period} and \ref{C24-period}).
Fig. \ref{fig:QBO-period} illustrates the latitude dependence of the QBO periods for Cycle 23 and Cycle 24 calculated using the observed shifts for the full frequency range (1900--4100\,$\mu$Hz). The dominant QBO period converges to values close to $\sim$3 years, particularly at latitudes above $20^\circ$ in both solar cycles.
This demonstrates that the underlying QBO periodicity is largely independent of frequency. The results obtained from the full frequency range provide the most physically robust characterisation of the QBO period. 
In general, Cycle 24 exhibits systematically longer QBO periods than Cycle 23 across most latitude bands (except for the 0-10 band), although the differences are smaller than the respective error bars. Cycle 23 presents a larger uncertainty in QBO behaviour than Cycle 24. 
This likely reflects the more complex temporal structure of Cycle 23 and the broader, less coherent wavelet power peaks seen in several latitude bands, which lead to larger period uncertainties.
We find that the QBO period is typically close to 3\,yr in both cycles within 1$\sigma$ uncertainties, except at low latitudes. This differs from \cite{2023ApJ...959...16J}, who reported a systematic difference, with ~2 yr in Cycle 23 and ~3 yr in Cycle 24. However, we note that the finding of a greater period in Cycle 24 compared to Cycle 23 is consistent with our results and the values are in agreement given the error bars. 
The discrepancy likely arises from differences in frequency range and data selection. While \cite{2023ApJ...959...16J} focused on modes from 1860–3450 $\upmu$Hz with 36-day non-overlapping data sets, our study analysed a broader range of 1900–4100 $\upmu$Hz using 108-day overlapping GONG data sets. 
Table \ref{table[nmodes]} demonstrates that there are of the order of hundreds of modes above $3450\,\rm\upmu Hz$ in each latitude band that will contribute. Such differences in frequency sensitivity may account for the prominence of the shorter periodicity in their analysis. Furthermore, the broad wavelet peaks in our results suggest that the quasi-periodic character of the QBO could accommodate reported values. 
Finally, spurious ~1 yr power from Earth’s orbital sampling may also complicate distinguishing between 2 and 3 yr components. Overall, we interpret our findings as consistent with a quasi-periodic modulation, with methodological differences explaining the cycle-to-cycle contrast seen in \cite{2023ApJ...959...16J}.

\section{QBO amplitude variations}\label{sec:qbo_amplitudes}
\begin{figure}
\begin{center}
\includegraphics[width=0.48\textwidth]{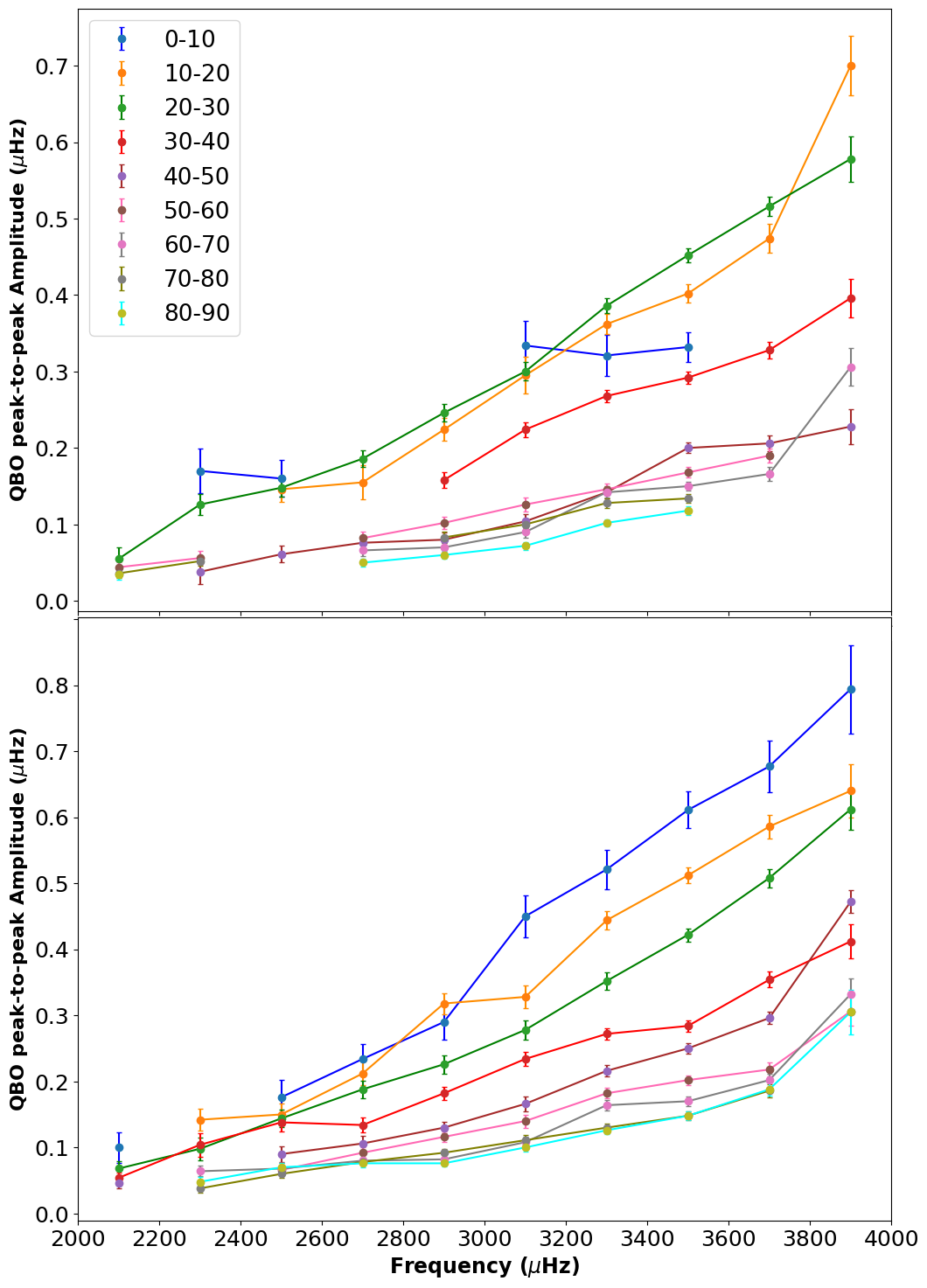}
\caption{QBO amplitude as a function of frequency and latitude (see legend) for Cycles 23 (lower panel) and 24 (upper).}
\label{fig:QBO-shift-amp}
\end{center}
\end{figure}

To investigate the variations in QBO amplitude, after removing the 11-year cycle (running-average curve) from the shift data (similar to the lower panel of Fig. \ref{fig:shift-smooth}), the maximum and minimum deviations in the residuals were calculated where extraction of QBO periodicity was feasible (as listed in Tables \ref{C23-period} and \ref{C24-period}). We define the QBO peak-to-peak amplitude as the difference between the maximum and minimum observed values. This definition provides the maximum observed QBO amplitude for each solar cycle and latitude band. 
However, we note that this value is an upper limit for the amplitude of QBO due to the noise present in the measurements. Fig. \ref{fig:QBO-shift-amp} demonstrates the variations in the QBO amplitude as a function of frequency for the latitude bands (only those for which a significant QBO periodicity was extracted according to Tables \ref{C23-period} and \ref{C24-period}). For both cycles, the QBO amplitude increases with frequency, indicating that higher-frequency modes tend to exhibit stronger QBO modulation. The $\theta<30$ degree latitude bands show the highest QBO amplitude, while the QBO amplitudes are relatively small at higher latitudes. The shape of the dependence of QBO amplitude on frequency is similar to that observed for the solar cycle \citep[e.g.][]{1990Natur.345..779L, 2001MNRAS.324..910C, 2017SoPh..292...67B} and is therefore symptomatic of a near-surface perturbation.

\begin{figure}
\begin{center}
\includegraphics[width=0.48\textwidth]{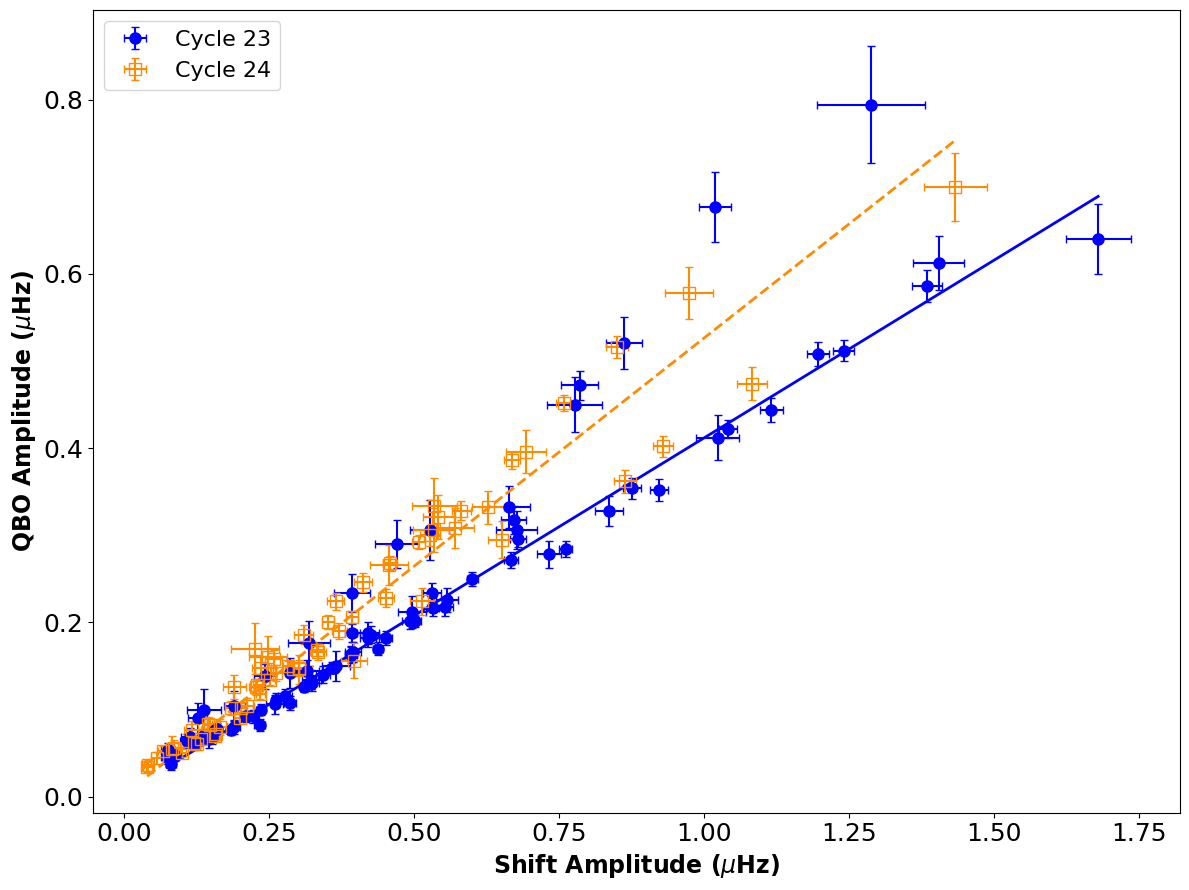}
\caption{Correlation between QBO amplitude and solar cycle shift peak-to-peak amplitude observed from the p-mode frequency shifts obtained when averaging over all combinations of frequency and latitude ranges where the QBO was found to be significant (as displayed in Tables \ref{C23-period} and \ref{C24-period}). The weighted least-squares linear fits are shown for Cycle 23 (blue solid) and Cycle 24 (dashed orange).}
\label{fig:QBO-shift-fit}
\end{center}
\end{figure}

To explore the amplitude trends in more detail, we plotted the QBO amplitude (Fig. \ref{fig:QBO-shift-amp}) as a function of the cycle amplitude (Fig. \ref{fig:amp-freq}) in both solar cycles. 
Fig. \ref{fig:QBO-shift-fit} reveals a clear positive correlation with Pearson correlation coefficients (r) $= 0.96$ for Cycle~23 and $r = 0.98$ for Cycle~24 between QBO and cycle peak-to-peak amplitude. 

To account for the varying number of contributing modes, we introduced statistical weights based on the number of common modes in each frequency range to the row total for each latitude band (see Table \ref{table[nmodes]}). To test the relation between QBO amplitude and solar-cycle amplitude, we applied a weighted least-squares linear fit, which found slopes of $0.41\pm0.01$ and $0.52\pm0.02$ for Cycles 23 and 24. 
The intercepts for these cycles were $0.004\pm0.006\,\upmu$Hz and $0.002\pm0.006\,\upmu$Hz, respectively. Also, the reduced chi-square values obtain 0.51 and 0.94 for Cycles 23 and 24, respectively, showing that the weighted linear model provides a statistically acceptable description of the data. The excess scatter in Cycle 23, in particular, may resemble hysteresis-like behaviour seen in other cycles and proxies \citep{1998A&A...329.1119J,2001SoPh..200....3T}. If the amplitude of the QBO were simply dependent on the strength of the magnetic field at any point, one might expect the gradients of the slopes plotted in Figure \ref{fig:QBO-shift-fit} to be the same. While magnetic field strength is clearly a dominant factor, the fact that the slopes are different in each cycle suggests that QBO amplitudes are not strictly proportional to cycle strength alone. This points to a partial decoupling between the QBO and the 11-year cycle, potentially reflecting additional dynamical processes in the solar interior.
The more pronounced QBO-to-cycle modulation seen in Cycle 24 likely reflects the overall weaker amplitude of that 11\,yr cycle.

\begin{figure*}
\begin{center}
\includegraphics[width=0.95\textwidth]{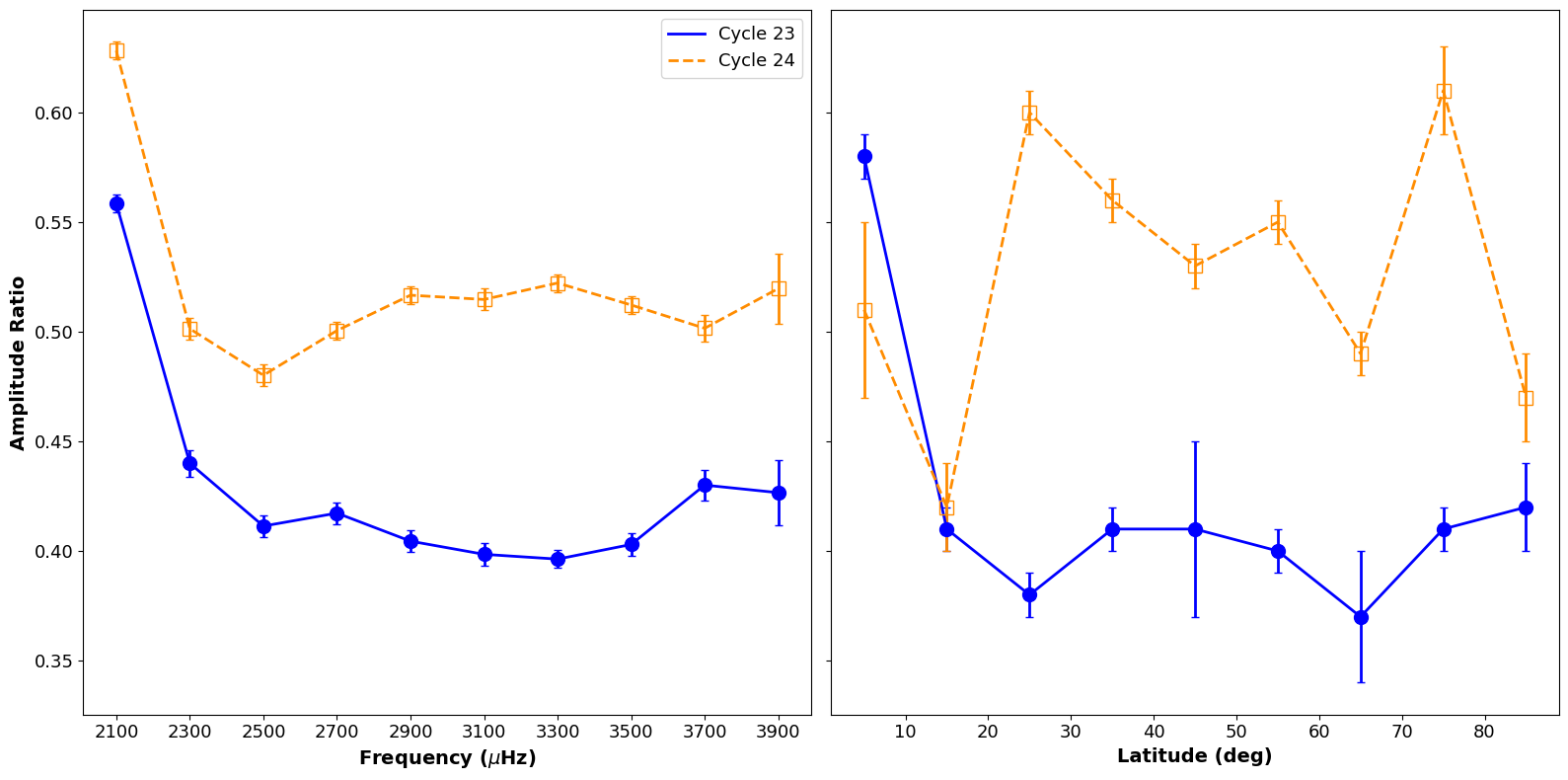}
\caption{Left panel: Amplitude ratio as a function of frequency for Cycles 23 (solid blue) and 24 (dashed orange line). The amplitudes were determined from shifts using the full solar disk. Right panel: QBO-to-cycle amplitude ratio variations with latitude for Cycles 23 (blue) and 24 (orange). The amplitudes were calculated using the full frequency range. 
}
\label{fig:QRatio}
\end{center}
\end{figure*}

Examination of the QBO-to-cycle amplitude ratio over the entire solar disk versus frequency is shown in the left panel of Fig. \ref{fig:QRatio} and indicates a consistent pattern observed in both Cycles 23 and 24. 
The cycle amplitude is defined as the peak-to-peak variation of the p-mode frequency shifts associated with the 11-year solar cycle, measured as the difference between the maximum frequency shift during cycle maximum and the minimum frequency shift observed at the subsequent cycle minimum. The QBO-to-cycle amplitude ratio is then defined as the ratio of the maximum QBO amplitude, derived from the residuals after removal of the smoothed 11-year trend, to this cycle amplitude.
The ratios are consistently higher in Cycle 24 compared to Cycle 23. In both cycles, the amplitude ratio declines up to about 2500 $\upmu$Hz, above which the amplitude ratios seem to stabilise.

The observed amplitude ratio is largest at low frequencies, which indicates that at low frequencies the amplitude of the QBO is relatively larger compared to the solar cycle than at higher frequencies. This is consistent with previous observations that suggest that the amplitude of the QBO decreases less rapidly at low frequencies compared to the 11\,yr solar cycle \citep[e.g.][]{2010ApJ...718L..19F, 2022MNRAS.515.2415M}.  However, at higher frequencies, the observed amplitude ratio becomes constant, suggesting that, at these higher frequencies, the frequency dependence is approximately the same for the QBO and the solar cycle. The higher amplitudes observed at the lowest-frequency band may partly reflect mode-selection effects, as this band contains fewer modes and therefore the results will be noisier. However, the deviation is far larger than the associated error bars. An alternative explanation is that this behaviour could be the result of the QBO magnetic perturbation extending over a larger range of depths than that associated with the solar cycle, since the upper turning points of low-frequency modes are deeper in the solar interior than those of high-frequency modes. We note that significant periodicities were detected in the frequency band $1900-2300\,\rm \upmu Hz$. The upper turning point of modes with frequencies of $2300\,\rm\upmu Hz$ is around $0.999R_\odot$ and therefore these results suggest that the magnetic perturbation extends to radii $<0.999R_\odot$.

We also investigated the amplitude ratio in the latitude bands for the two cycles by using the full frequency range (right panel of Fig. \ref{fig:QRatio}). For Cycle 23, the amplitude ratio remains broadly constant above 20 degrees, consistent with the same latitudinal distribution for the QBO as for the 11\,yr cycle. In Cycle 24, modest variations are present across latitude bands, but these lie within the associated uncertainties and are therefore not statistically significant. Overall, both cycles display a broadly similar latitudinal distribution of the amplitude ratio, although Cycle 24 shows greater scatter.
Discrepancies at the lower latitude range may be due to the fact that the statistical significance of the QBO is weaker here.

\begin{figure}
\begin{center}
\includegraphics[width=0.48\textwidth]{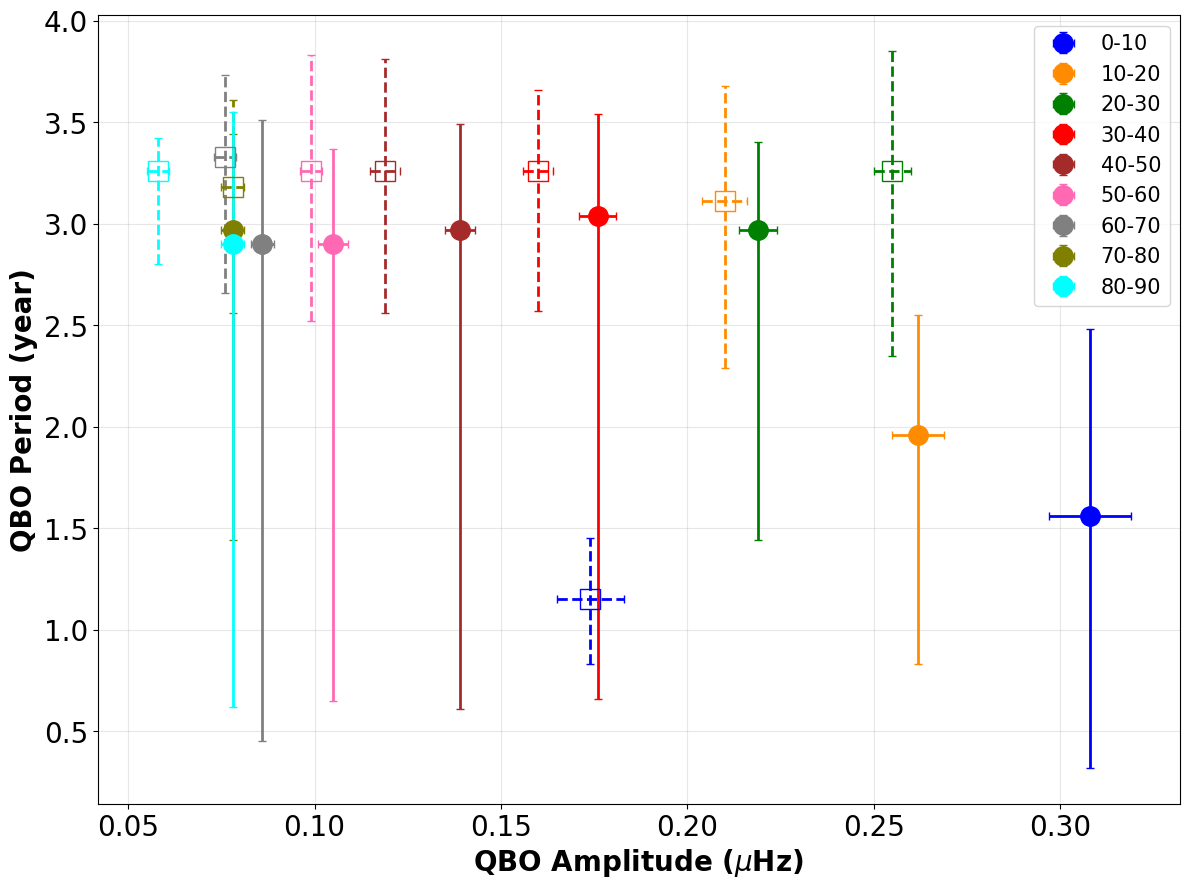}
\caption{QBO period as a function of QBO amplitude for Cycle 23 (filled circles) and 24 (open squares with dashed line error bar) in latitude bands and using modes from the entire frequency range. Error bars in both dimensions reflect the uncertainties in measured QBO amplitudes and periods.}
\label{fig:amp-period}
\end{center}
\end{figure}

We obtained the QBO amplitudes using shifts across various latitude bands within the full frequency range and combined them with QBO periods (see Fig. \ref{fig:QBO-period}). In general, Cycle 23 exhibits higher QBO amplitudes than Cycle 24 and shorter periods than Cycle 24, although again we note that for the periods the difference is smaller than the associated uncertainties (Fig. \ref{fig:amp-period}). 
On the other hand, Figure \ref{fig:amp-period} shows almost no dependence of the QBO period upon amplitude across latitude bands for both cycles. This behaviour is consistent with a linear oscillator interpretation of the QBO phenomenon.

\section{Conclusions}
We have used helioseismic p-mode frequency shifts, observed by GONG and grouped by their frequencies and latitudinal dependences, to study the spatial and temporal characteristics of the quasi-biennial oscillation (QBO) through cycles 23 and 24. We have used wavelets to detect and characterise statistically significant periodicities, while the amplitudes of statistically significant QBOs were determined by the amplitudes of the residuals once a smoothed trend (representing the 11-yr solar cycle) was subtracted. These amplitudes were compared to the amplitudes of the solar cycle, as seen in the p-mode frequency shifts, and determined by the difference between the maximum frequency shift and the minimum frequency shift (observed in the cycle minimum that followed the respective maximum). The key findings are as follows.
\begin{itemize}

\item
QBO periodicities exhibit some latitudinal dependence, with shorter periods typically found at latitudes less than 20 degrees. However, no latitudinal dependence in QBO period was observed at latitudes above 20 degrees. Notably, the $0\le \theta < 10$ degree band displayed irregular and less persistent signals. This anomaly highlights the complex nature of solar cycle progression at low latitudes \citep{2020LRSP...17....4C}.

\item
Cycle 24 tended to exhibit slightly longer QBO periods compared to Cycle 23, although the difference was smaller than the associated uncertainties. For latitudes greater than 20 degrees, the QBO periods were approximately constant with latitude and had values of around 3\,yrs. 
The relatively wide range of inferred QBO periods reflects both the quasi-periodic nature of the signal and methodological limitations. In particular, broader 
wavelet power peaks and temporal intermittency lead to larger uncertainties in period determination, especially in Cycle~23. The larger uncertainties and higher reduced $\chi^2$ values found in Cycle 23 
likely arise from its more complex temporal evolution and increased variability in the residual frequency shifts compared to the weaker and more uniform Cycle 24.

\item The amplitude of the QBO increased with frequency. This implies that higher-frequency modes, which are more sensitive to shallower subsurface layers, respond more strongly to QBO-related perturbations. However, we did not perform structural inversions to determine the exact depth of this perturbation. Therefore, our results only provide qualitative constraints on its radial location. However, the behaviour is nevertheless consistent with previous helioseismic studies suggesting that QBO-related perturbations may extend deeper than those associated with the 11-yr cycle \citep{2010ApJ...718L..19F, 2022MNRAS.515.2415M}. The QBO was present at all latitudes, in agreement with previous studies \citep[e.g.][]{2021ApJ...920...49I}. However, higher amplitude QBOs were observed for $\theta< 30$\,degrees compared to $\theta > 30$\,degrees, which likely reflects the fact that the p-mode perturbations are often associated with the active latitudes where strong photospheric magnetic fields are observed \citep{2002ApJ...580.1172H}.

\item The ratio of the QBO amplitude to the cycle amplitude is higher in Cycle 24 than Cycle 23. 
Across latitude bands, the amplitude ratio in Cycle 23 remains broadly uniform above 20 degrees, whereas Cycle 24 exhibits modest variations. A linear correlation was found between QBO amplitude and solar cycle amplitude. However, the gradients of the slopes were significantly different, with the steeper gradient found in Cycle 24. 
The significantly different slopes indicate that QBO amplitudes are not wholly governed by the 11-yr cycle strength, pointing to a partial decoupling of the two phenomena. With only two cycles of helioseismic data at our disposal, it is difficult to know how significant this apparent consistency in QBO amplitude from one cycle to the next is, thus motivating a more long-term study, which spans far longer epochs. If real, however, dynamo models may be constrained by the need to ensure QBO amplitudes are not determined by or are only partially modulated by cycle amplitudes. This observation also raises the possibility that a different method of normalising by the cycle amplitude may reveal the QBO at times of cycle minima, as seen recently in the polar faculae observations \citep{2020MNRAS.494.4930D} and helioseismic solar pseudomode observations \citep{2024SoPh..299..134M}.

\item The observed QBO period was found to not depend on the observed QBO amplitude. This was the case for all latitude bands considered. Such behaviour (constant period despite changes in amplitude) is consistent with a linear oscillation regime, which again provides additional constraints for models.

\end{itemize}

\section*{Data Availability}
All data used in this study are publicly available from the Global Oscillations Network Group (GONG) and National Solar Observatory (NSO) websites. The wavelet analysis figures used within this article are available as a \href{https://drive.google.com/drive/folders/1Bvxxafb5LkuFrjSERpAnzKjXj4ozzA1Q} {supplementary appendix}.

\section*{Acknowledgement}
Authors acknowledge support from the Science and Technology Facilities Council (STFC) grant No. ST/X000915/1 and ST/T000252/1. DYK also acknowledge the Latvian Council of Science Project No.~lzp-2024/1-0023 and the UKRI Stephen Hawking Fellowship EP/Z535473/1. QBO analysis code is available at {\url{https://github.com/TishtryaMehta/QBO_evolution}}.
This work utilises data from the National Solar Observatory Integrated Synoptic Program, which is operated by the Association of Universities for Research in Astronomy, under a cooperative agreement with the National Science Foundation and with additional financial support from the National Oceanic and Atmospheric Administration, the National Aeronautics and Space Administration, and the United States Air Force. The Big Bear Solar Observatory, High Altitude Observatory, Learmonth Solar Observatory, Udaipur Solar Observatory, Instituto de Astrofísica de Canarias, and Cerro Tololo Interamerican Observatory host the GONG network of instruments. 



\bibliographystyle{mnras}
\bibliography{ref} 

\appendix
\section{Additional Tables}\label{sec:app_tables}
\begin{landscape}
\begin{table}
  \caption{Number of common modes for each latitude and frequency range considered.}
  \label{table[nmodes]}
       \begin{tabular} {c| c| c| c| c |c |c |c |c |c| c | c }
\hline
Latitude band &
$1900-2300$ &
$2100-2500$ &
$2300-2700$ &
$2500-2900$ &
$2700-3100$ &
$2900-3300$ & 
$3100-3500$ &
$3300-3700$ &
$3500-3900$ &
$3700-4100$ &
$1900-4100$
 \\
  \hline 
   $0-10$ & $13$ & $11$ & $17$ & $24$ & $16$ & $15$ & $58$ & $103$ & $83$ & $27$ & $189$ \\
  $10-20$ & $43$ & $35$ & $37$ & $39$ & $45$ & $42$ & $142$ & $252$ & $197$ & $79$ & $475$ \\
   $20-30$ & $55$ & $46$ & $66$ & $77$ & $78$ & $84$ &  $240$ & $434$ & $365$ & $132$ & $813$ \\
   $30-40$ & $85$ & $57$ & $75$ & $89$ & $96$ & $104$ & $317$ & $595$ & $504$ & $190$ & $1098$ \\
   $40-50$ & $103$ & $75$ & $87$ & $108$ & $128$ & $127$ & $362$ & $742$ & $651$ & $222$ & $1343$ \\
   $50-60$ & $141$ & $115$ & $130$ & $148$ & $160$ & $147$ & $413$ & $843$ & $683$ & $194$ & $1537$ \\
    $60-70$ & $191$ & $155$ & $180$ & $233$ & $256$ & $217$ & $479$ & $917$ & $704$ & $179$ & $1814$ \\
   $70-80$ & $261$ & $232$ & $287$ &  $344$ & $379$ & $359$ & $632$ & $1015$ & $734$ & $198$ & $2300$ \\
    $80-90$ & $248$ & $271$ & $324$ & $391$ & $446$ & $419$ & $658$ & $1027$ & $754$ & 
    $202$ & $2433$ \\
    \hline
    $0-90$ & $1140$ & $997$ & $1201$ & $1453$ & $1604$ & $1514$ & $3301$ & $5928$ & $4675$ & $1423$ & $12002$\\
  \hline
  \hline
   \end{tabular}
 \\ 
\end{table}
\end{landscape}

\begin{landscape}
\begin{table}
 \begin{footnotesize}
 \setlength\tabcolsep{3pt}
 \renewcommand{\arraystretch}{1.6}
  \caption{Helioseismic QBO periods (years) in latitude bands for cycle 23. For some bands, no periodicity was found with a confidence higher than 0.95, in which case they are marked with "-".}
  \label{C23-period}
       \begin{tabular} {c| c| c| c| c |c |c |c |c |c|c|c}
\hline
Latitude band &
$1900-2300$ &
$2100-2500$ &
$2300-2700$ &
$2500-2900$ &
$2700-3100$ &
$2900-3300$ & 
$3100-3500$ &
$3300-3700$ &
$3500-3900$ &
$3700-4100$ &
$1900-4100$
 \\
  \hline 
   $0-10$ & 
   $3.83^{+0.32}_{-0.38}$ & 
           -              & 
   $1.21^{+0.17}_{-0.22}$ & 
   $1.63^{+0.38}_{-0.31}$ & 
   $1.67^{+0.46}_{-0.37}$ & 
   $1.56^{+1.56}_{-1.04}$ &  
   $1.52^{+1.93}_{-1.01}$ &
   $1.52^{+1.93}_{-1.02}$ &
   $1.52^{+0.74}_{-1.03}$ &
   $3.49^{+0.29}_{-0.40}$ &
   $1.56^{0.92}_{-1.24}$\\
    $10-20$ &
            -             & 
            
     $2.10^{+0.32}_{-1.22}$ & $1.75^{+0.47}_{-0.87}$ & $1.96^{+0.63}_{-0.66}$ & $2.15^{+0.59}_{-1.34}$ & $2.10^{+0.50}_{-1.25}$ &  $1.96^{+0.64}_{-1.16}$ & $1.96^{+0.58}_{-1.19}$ &
     $1.96^{+0.49}_{-0.53}$ &
     $2.10^{+0.34}_{-0.73}$ &
     $1.96^{+0.59}_{-1.13}$\\
   $20-30$ &
      $3.18^{+0.60}_{-0.48}$ &        -               & $3.04^{+0.46}_{-0.93}$ & $2.97^{+0.56}_{-1.61}$ & $2.97^{+0.38}_{-1.65}$ & $2.97^{+0.35}_{-1.61}$ &  $2.84^{+0.53}_{-1.43}$ &
      $2.90^{+0.51}_{-1.55}$ &
      $1.75^{+1.55}_{-0.75}$ & $1.71^{+0.49}_{-0.70} $ &
      $2.97^{+0.43}_{-1.53}$\\
   $30-40$ &
      $1.52^{+0.20}_{-0.23}$ & 
      $1.10^{+0.76}_{-0.29}$ &
      $1.63^{+0.46}_{-1.16}$ & $1.67^{+0.54}_{-0.67}$ & $3.04^{+0.57}_{-1.71}$ & $3.04^{+0.45}_{-2.40}$ & $3.04^{+0.39}_{-2.35}$ &   $3.04^{+0.53}_{-2.35}$ &  $3.04^{+0.60}_{-2.33}$  & $3.04^{+0.42}_{-0.57}$ & $3.04^{+0.50}_{-2.38}$\\
   $40-50$ &
      $1.32^{+0.22}_{-0.40}$ &                      -                  &   $1.21^{+0.56}_{-0.27}$   & $2.71^{+0.63}_{-1.24}$ & $2.84^{+0.57}_{-1.39}$ & $2.90^{+0.54}_{-2.29}$ &  $2.97^{+0.53}_{-2.32}$ & $2.90^{+0.57}_{-2.28}$ & $2.90^{+0.55}_{-1.58}$ &
      $2.97^{+0.56}_{-0.53}$ & $2.97^{+0.52}_{-2.36}$  \\
   $50-60$ &
                -             &
                -             & $3.11^{+0.25}_{-0.63}$ & $3.11^{+0.26}_{-0.74}$ & $2.97^{+0.30}_{-0.68}$ & $2.90^{+0.57}_{-2.19}$ & $2.97^{+0.48}_{-2.39}$ &  $2.90^{+0.51}_{-2.31}$ & $2.77^{+0.63}_{-2.05}$ & $2.71^{+0.70}_{-1.01}$ & $2.97^{+0.47}_{-2.25}$\\
    $60-70$ &
                -             &
      $1.63^{+1.23}_{-0.45}$ & $2.25^{+0.61}_{-0.84}$ & $2.59^{+0.58}_{-1.07}$ & $2.97^{+0.50}_{-1.18}$ & $3.04^{+0.52}_{-1.31}$ & 
      $2.90^{+0.55}_{-2.47}$  &
      $2.84^{+0.63}_{-2.26}$ & $2.84^{+0.55}_{-2.17}$ & $1.79^{+0.24}_{-0.44}$ & $2.90^{+0.61}_{-2.45}$\\
   $70-80$ & 
                -             &
      $2.53^{+0.14}_{-0.21}$ &
      $1.56^{+0.34}_{-0.42}$ &  $3.04^{+0.36}_{-0.54}$ & $2.97^{+0.54}_{-2.14}$ & 
      $2.97^{+0.53}_{-2.33}$ &
      $2.97^{+0.51}_{-2.18}$ &  
      $2.97^{+0.49}_{-2.21}$ &
      $2.90^{+0.52}_{-1.90}$ &
                -            &
      $2.97^{+0.47}_{-1.53}$ \\
    $80-90$ &
                -             &
      $3.11^{+0.55}_{-0.66}$ & $2.77^{+0.75}_{-0.94}$ & $2.90^{+0.38}_{-1.17}$ & $2.97^{+0.36}_{-1.54}$ & $2.77^{+0.63}_{-1.32}$ & 
      $2.77^{+0.65}_{-1.37}$  &  
      $2.97^{+0.49}_{-2.30}$ & $2.97^{+0.50}_{-2.25}$ & 
      $2.97^{+0.53}_{-0.42}$  &
      $2.90^{+0.65}_{-2.28}$\\
  \hline 
     $0-90$ &
      $3.18^{+0.05}_{-0.18}$ &
      $3.11^{+0.47}_{-0.86}$ & $2.97^{+0.53}_{-1.59}$ & $2.90^{+0.56}_{-2.24}$ & $2.90^{+0.60}_{-2.24}$ & $2.90^{+0.54}_{-2.29}$ & 
      $2.90^{+0.53}_{-2.26}$  &  
      $2.84^{+0.59}_{-2.20}$ & $2.84^{+0.59}_{-2.21}$ & 
      $2.90^{+0.43}_{-2.42}$  &
      $2.90^{+0.55}_{-2.26}$\\
  \hline
  \hline
   \end{tabular}
 \\ 
\end{footnotesize}
\end{table}
\end{landscape}

\begin{landscape}
\begin{table}
 \begin{footnotesize}
 \setlength\tabcolsep{3pt}
 \renewcommand{\arraystretch}{1.6}
  \caption{Helioseismic QBO periods in latitude bands for cycle 24.}
  \label{C24-period}
       \begin{tabular} {c| c| c| c| c |c |c |c |c |c| c| c}
\hline
Latitude band &
$1900-2300$ &
$2100-2500$ &
$2300-2700$ &
$2500-2900$ &
$2700-3100$ &
$2900-3300$ & 
$3100-3500$ &
$3300-3700$ &
$3500-3900$ &
$3700-4100$ &
$1900-4100$\\
  \hline 
   $0-10$ & 
          -                & 
   $2.90^{+0.34}_{-0.49}$ & 
   $3.04^{+0.33}_{-0.42}$ & 
            -              & 
            -              & 
   $1.15^{+0.33}_{-0.33}$ &  
   $1.21^{+0.37}_{-0.47}$ &
   $1.18^{+0.31}_{-0.44}$ &
            -              &
            -              &
    $1.15^{+0.30}_{-0.32}$\\
   $10-20$ &
            -               & 
            -              & 
    $2.97^{+0.05}_{-0.06}$ & 
    $3.11^{+0.04}_{-0.06}$ & 
    $3.11^{+0.34}_{-0.50}$ & 
    $3.11^{+0.47}_{-0.60}$ &  
    $2.97^{+0.68}_{-0.73}$ & 
    $2.97^{+0.74}_{-0.75}$ &
    $3.11^{+0.69}_{-0.84}$ &
    $3.04^{+0.77}_{-0.72}$ &
    $3.11^{+0.57}_{-0.82}$\\
   $20-30$ &
    $3.11^{+0.22}_{-0.25}$ & 
    $3.04^{+0.71}_{-0.78}$ &
    $3.18^{+0.52}_{-0.63}$ & 
    $3.26^{+0.34}_{-0.52}$ &
    $3.26^{+0.48}_{-0.75}$ & 
    $3.26^{+0.48}_{-0.70}$ & 
    $3.18^{+0.58}_{-0.79}$ &
    $3.18^{+0.59}_{-0.89}$ &
    $3.18^{+0.52}_{-0.88}$ & 
    $3.18^{+0.47}_{-0.72}$ &
    $3.26^{+0.59}_{-0.91}$ \\
   $30-40$ &
            -               & 
            -               &
            -              &
            -               &
    $3.18^{+0.29}_{-0.52}$ & 
    $3.18^{+0.51}_{-0.70}$ & 
    $3.26^{+0.47}_{-0.74}$ & 
    $3.26^{+0.42}_{-0.73}$ & 
    $3.26^{+0.36}_{-0.63}$ & 
    $3.41^{+0.43}_{-0.56}$ &
    $3.26^{+0.40}_{-0.69}$\\
   $40-50$ &
            -              & 
    $2.71^{+0.43}_{-0.39}$ &
    $3.18^{+0.29}_{-0.44}$ & 
    $3.26^{+0.36}_{-0.47}$ &
    $3.18^{+0.45}_{-0.59}$ & 
    $3.33^{+0.49}_{-0.73}$ &  
    $3.33^{+0.58}_{-0.81}$ & 
    $3.26^{+0.46}_{-0.68}$ & 
    $3.26^{+0.46}_{-0.77}$ &
    $3.26^{+0.36}_{-0.41}$ &
    $3.26^{+0.55}_{-0.76}$ \\
   $50-60$ &
      $3.33^{+0.19}_{-0.38}$ &
      $3.18^{+0.36}_{-0.31}$ & 
               -             & 
      $3.04^{+0.41}_{-0.47}$ & 
      $3.11^{+0.53}_{-0.69}$ &
      $3.26^{+0.34}_{-0.58}$ &
      $3.18^{+0.50}_{-0.60}$ &     
      $3.26^{+0.48}_{-0.66}$ & 
      $3.41^{+0.11}_{-0.35}$ &  
               -             &
      $3.26^{+0.57}_{-0.74}$\\
    $60-70$ &
             -              &                   
             -              &
             -              & 
    $3.26^{+0.43}_{-0.42}$ & 
    $3.26^{+0.39}_{-0.49}$ & 
    $3.26^{+0.28}_{-0.55}$ &  
    $3.18^{+0.33}_{-0.59}$ & 
    $3.33^{+0.51}_{-0.80}$ & 
    $3.49^{+0.43}_{-0.72}$ &
    $2.90^{+0.57}_{-0.50} $ &
    $3.33^{+0.40}_{-0.67}$\\
   $70-80$ & 
      $3.04^{+0.51}_{-0.44}$ &
      $3.33^{+0.06}_{-0.04}$ &
             -                &
             -                &
      $3.11^{+0.14}_{-0.36}$ & 
      $3.11^{+0.34}_{-0.60}$ &
      $3.33^{+0.35}_{-0.77}$ &  
      $3.41^{+0.18}_{-0.85}$ &
             -               &
             -               & 
      $3.18^{+0.43}_{-0.62}$ \\
    $80-90$ &
      $1.21^{+0.26}_{-0.19}$ &
             -               & 
             -                &
      $3.18^{+0.25}_{-0.34}$ & 
      $3.18^{+0.38}_{-0.50}$ &
      $3.33^{+0.24}_{-0.60}$ &
      $3.33^{+0.10}_{-0.42}$ &     
      $3.18^{+0.08}_{-0.40}$ & 
              -               &  
              -               &
      $3.26^{+0.16}_{-0.46}$\\
  \hline 
     $0-90$ &
      $3.18^{+0.05}_{-0.06}$ &
      $3.18^{+0.01}_{-0.12}$ & $3.18^{+0.40}_{-0.54}$  & $3.18^{+0.45}_{-0.60}$  & $3.18^{+0.53}_{-0.65}$  & $3.26^{+0.49}_{-0.75}$  & 
      $3.26^{+0.53}_{-0.81}$  &  
      $3.26^{+0.51}_{-0.80}$& $3.26^{+0.46}_{-0.77}$ & 
      $3.33^{+0.49}_{-0.79}$  &
      $3.26^{+0.53}_{-0.78}$\\
  \hline
  \hline
   \end{tabular}
 \\ 
\end{footnotesize}
\end{table}
\end{landscape}


\section{Additional Figures}\label{sec:app_figures}
\begin{figure*}
\begin{center}
\includegraphics[width=0.48\textwidth]{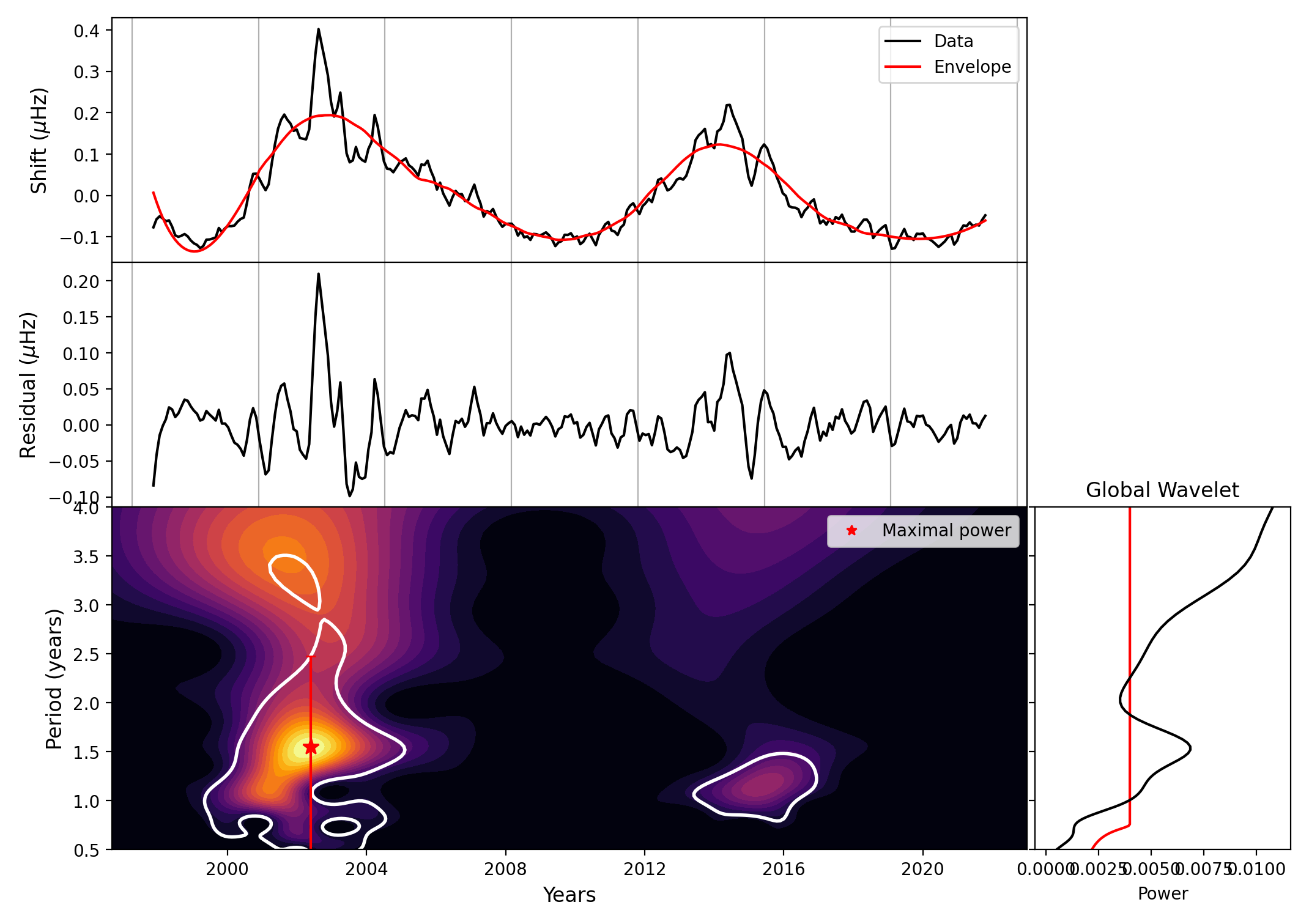}
\caption{Wavelet power spectra for the latitude band 0–10 across the full frequency range (1900–4100 $\mu$Hz)}
\label{fig:1900_4100-0_10}
\end{center}
\end{figure*}

\begin{figure}
\begin{center}
\includegraphics[width=0.48\textwidth]{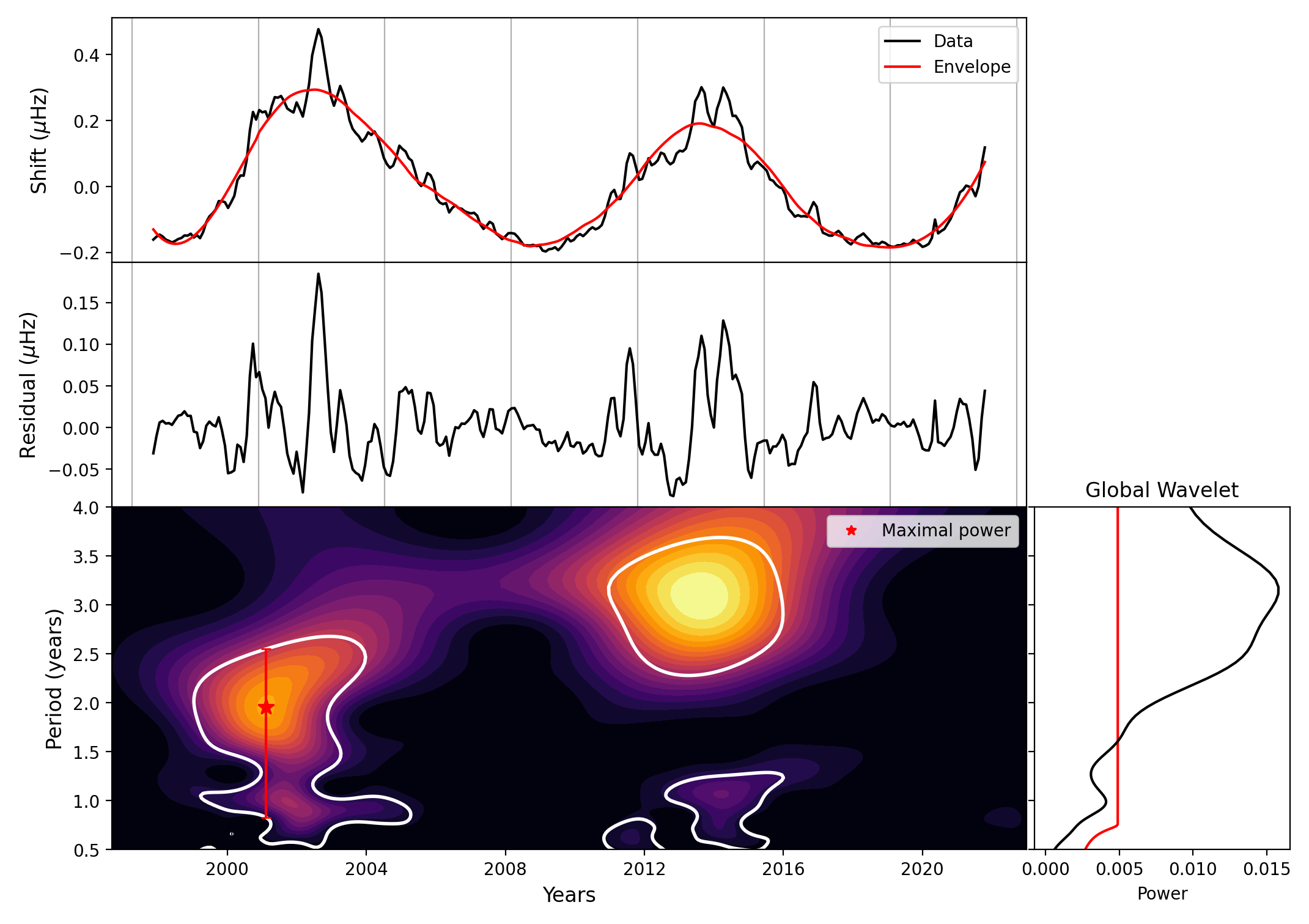}
\caption{As in Fig.~\ref{fig:1900_4100-0_10}, but for latitude band 10–20.}
\label{fig:1900_4100-10_20}
\end{center}
\end{figure}

\begin{figure}
\begin{center}
\includegraphics[width=0.48\textwidth]{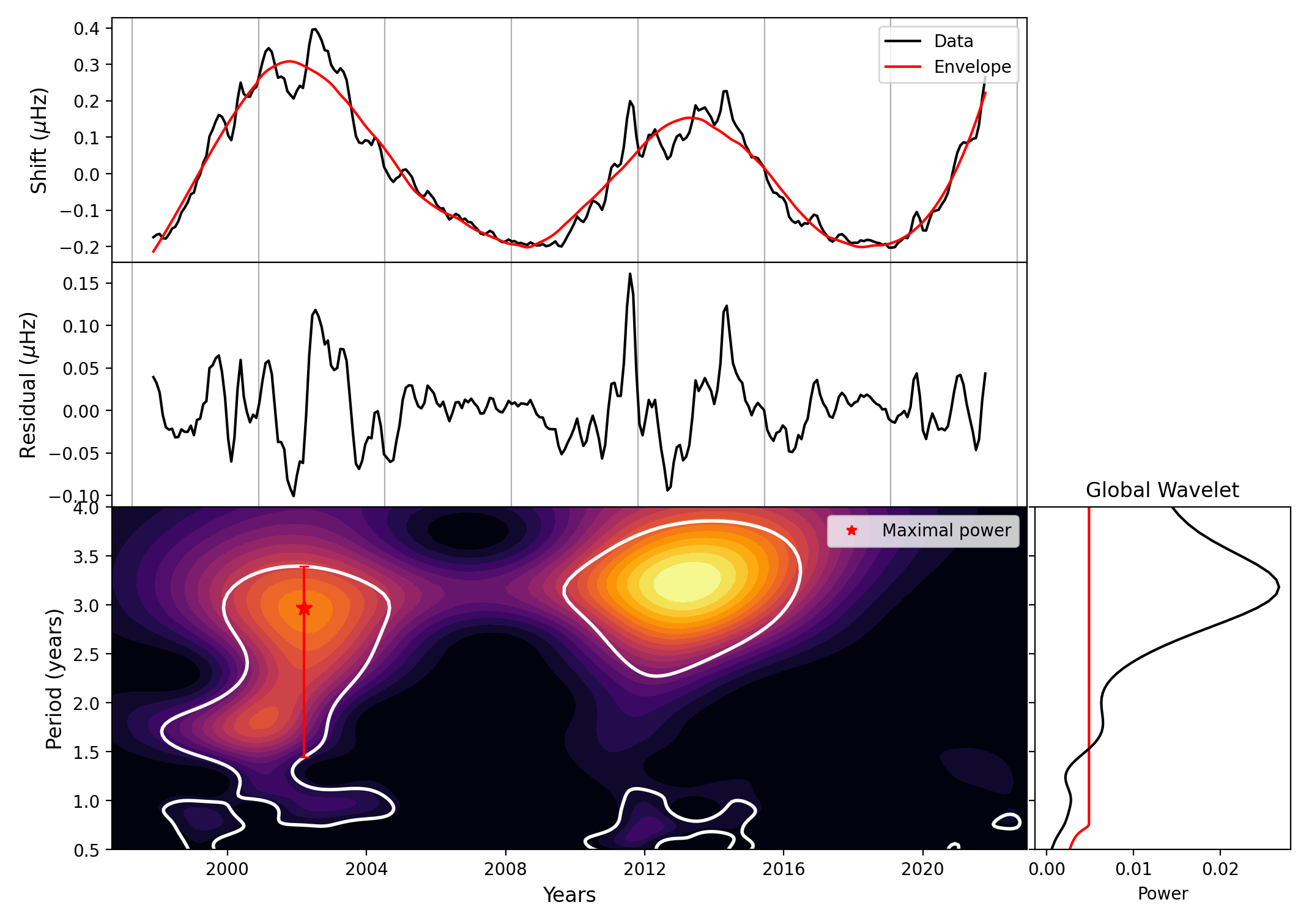}
\caption{As in Fig.~\ref{fig:1900_4100-0_10}, but for latitude band 20–30.}
\label{fig:1900_4100-20_30}
\end{center}
\end{figure}

\begin{figure}
\begin{center}
\includegraphics[width=0.48\textwidth]{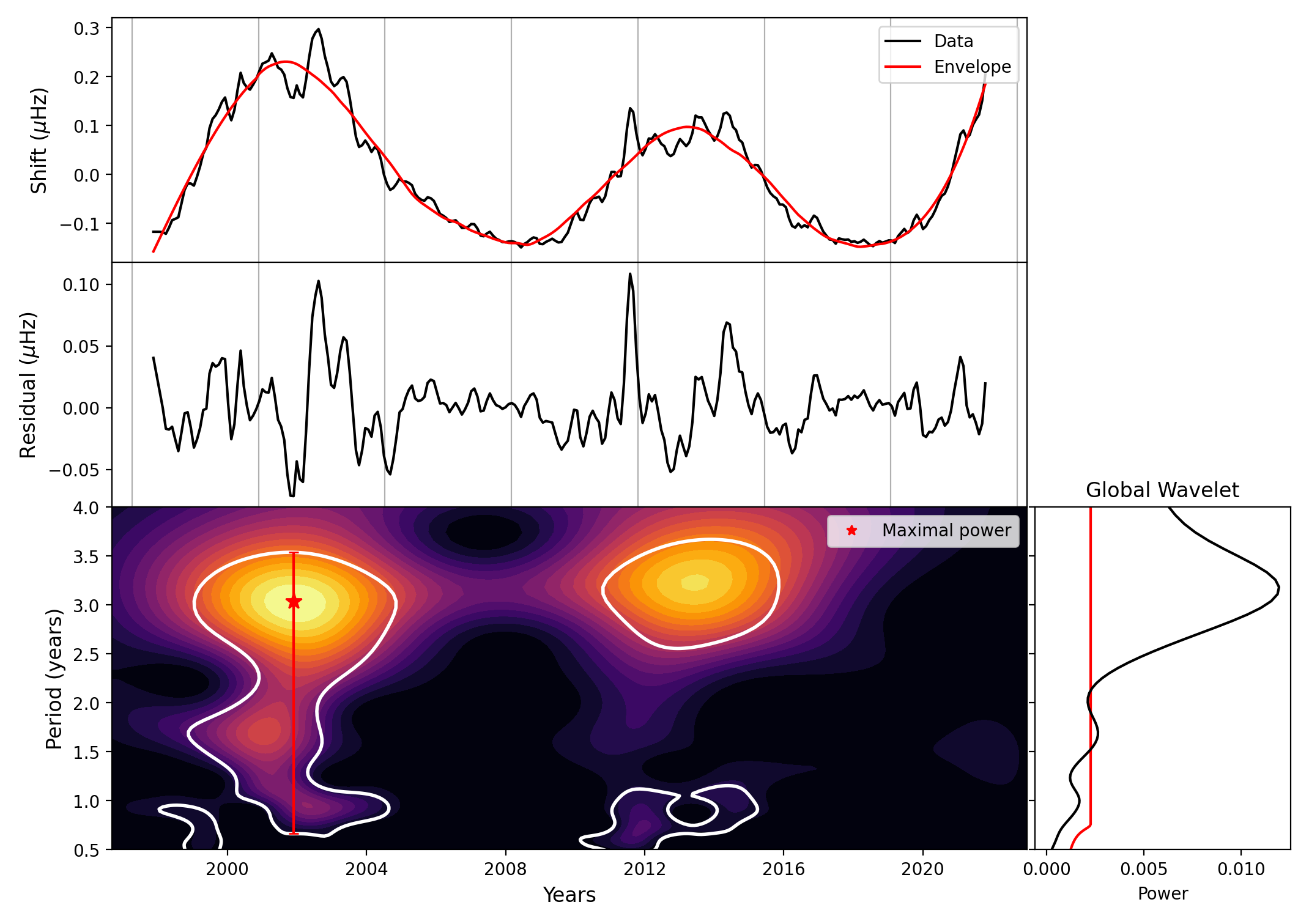}
\caption{Wavelet power spectra for latitude band 30–40.}
\label{fig:1900_4100-30_40}
\end{center}
\end{figure}

\begin{figure}
\begin{center}
\includegraphics[width=0.48\textwidth]{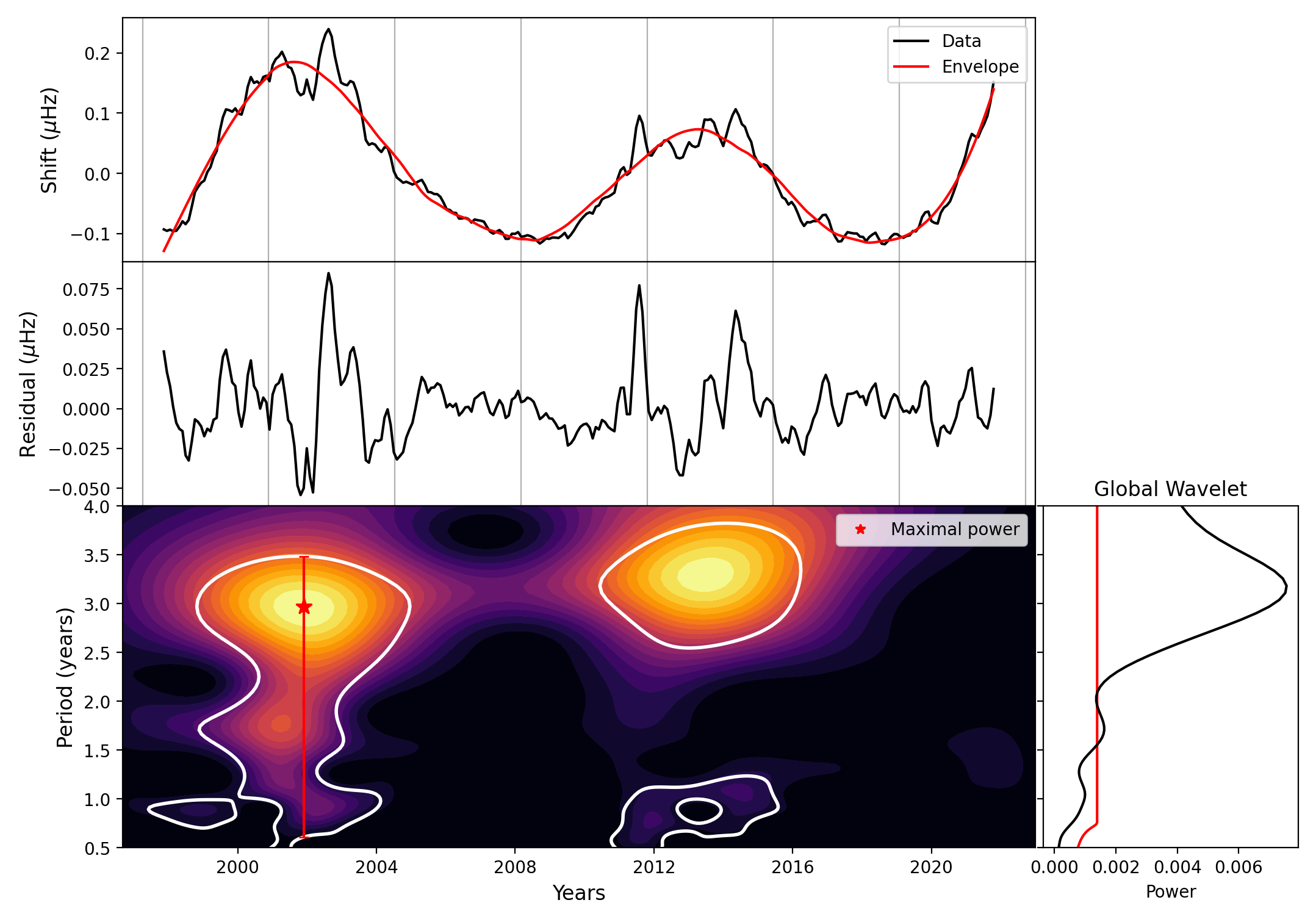}
\caption{Wavelet power spectra for latitude band 40–50}
\label{fig:1900_4100-40_50}
\end{center}
\end{figure}

\begin{figure}
\begin{center}
\includegraphics[width=0.48\textwidth]{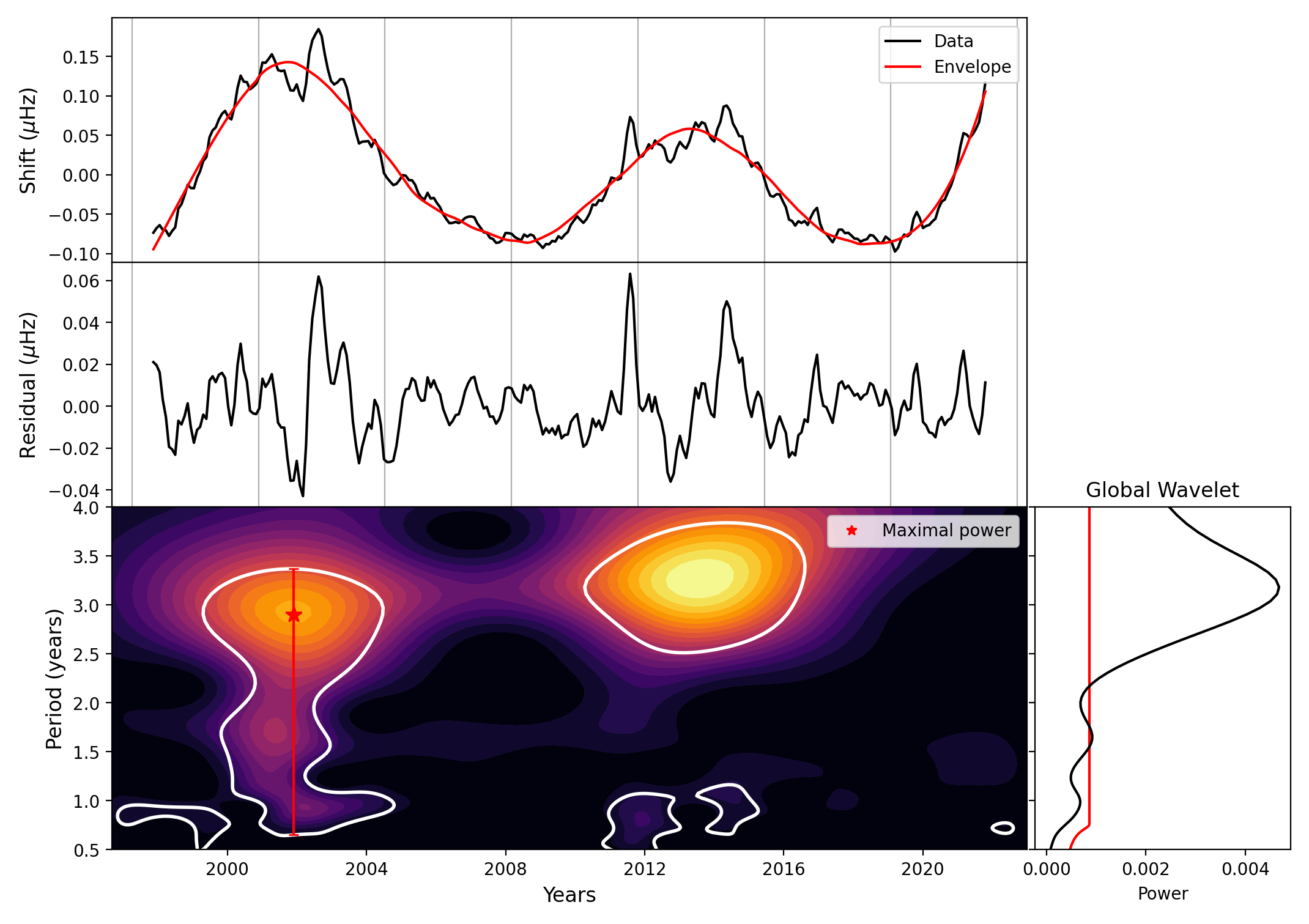}
\caption{Wavelet power spectra for latitude band 50–60}
\label{fig:1900_4100-50_60}
\end{center}
\end{figure}

\begin{figure}
\begin{center}
\includegraphics[width=0.48\textwidth]{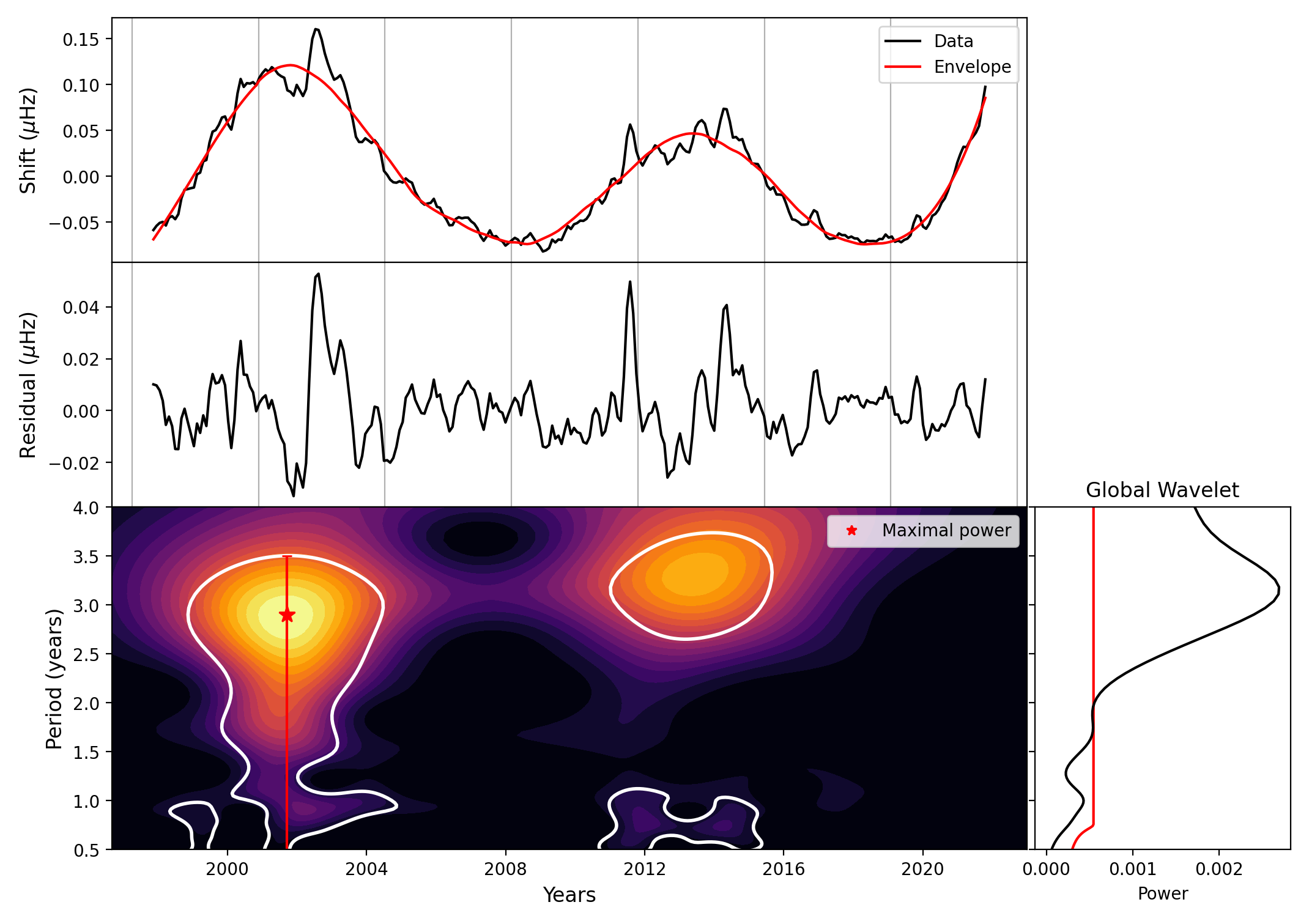}
\caption{Wavelet power spectra for latitude band 60–70}
\label{fig:1900_4100-60_70}
\end{center}
\end{figure}

\begin{figure}
\begin{center}
\includegraphics[width=0.48\textwidth]{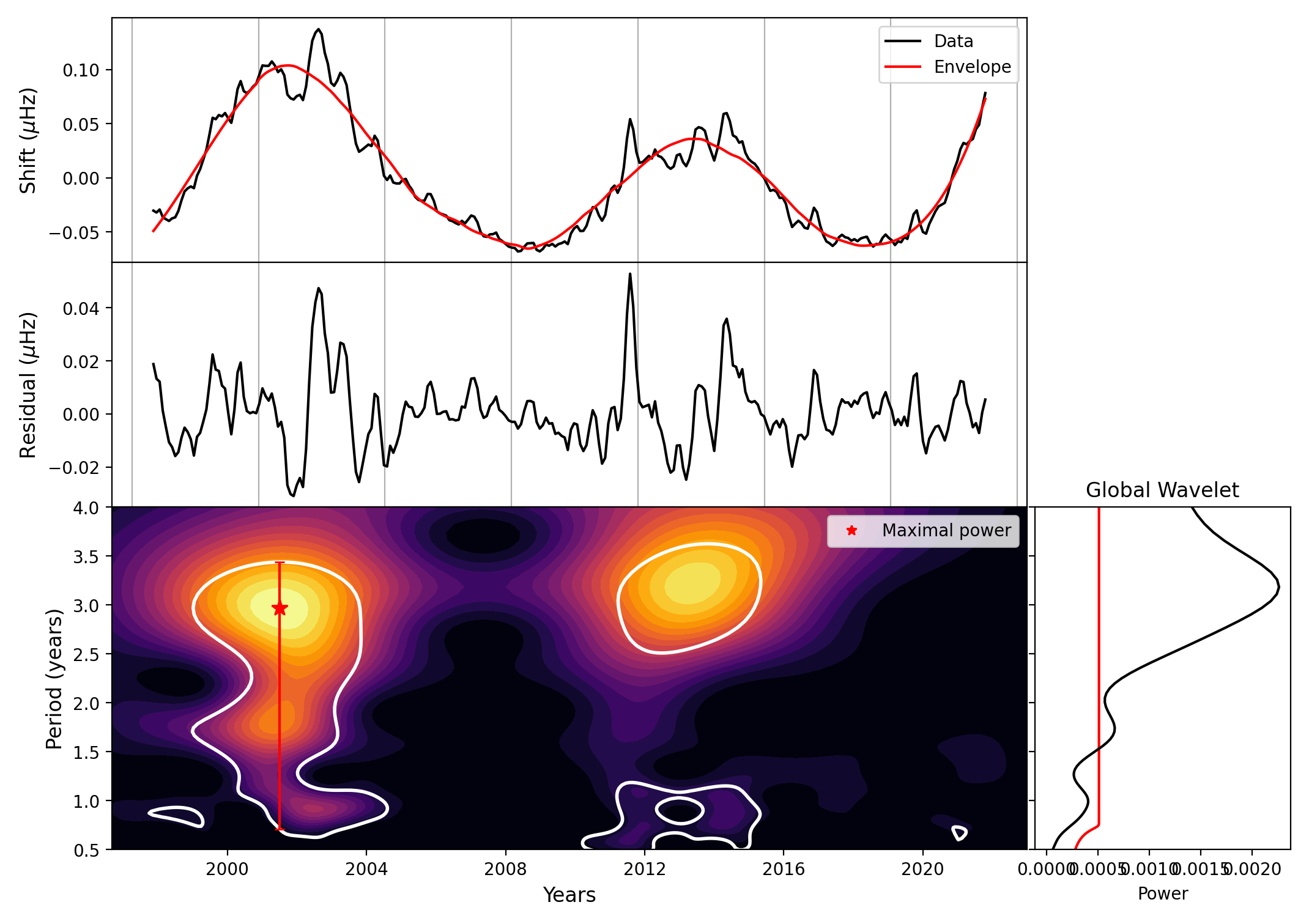}
\caption{Wavelet power spectra for latitude band 70–80}
\label{fig:1900_4100-70_80}
\end{center}
\end{figure}

\begin{figure}
\begin{center}
\includegraphics[width=0.48\textwidth]{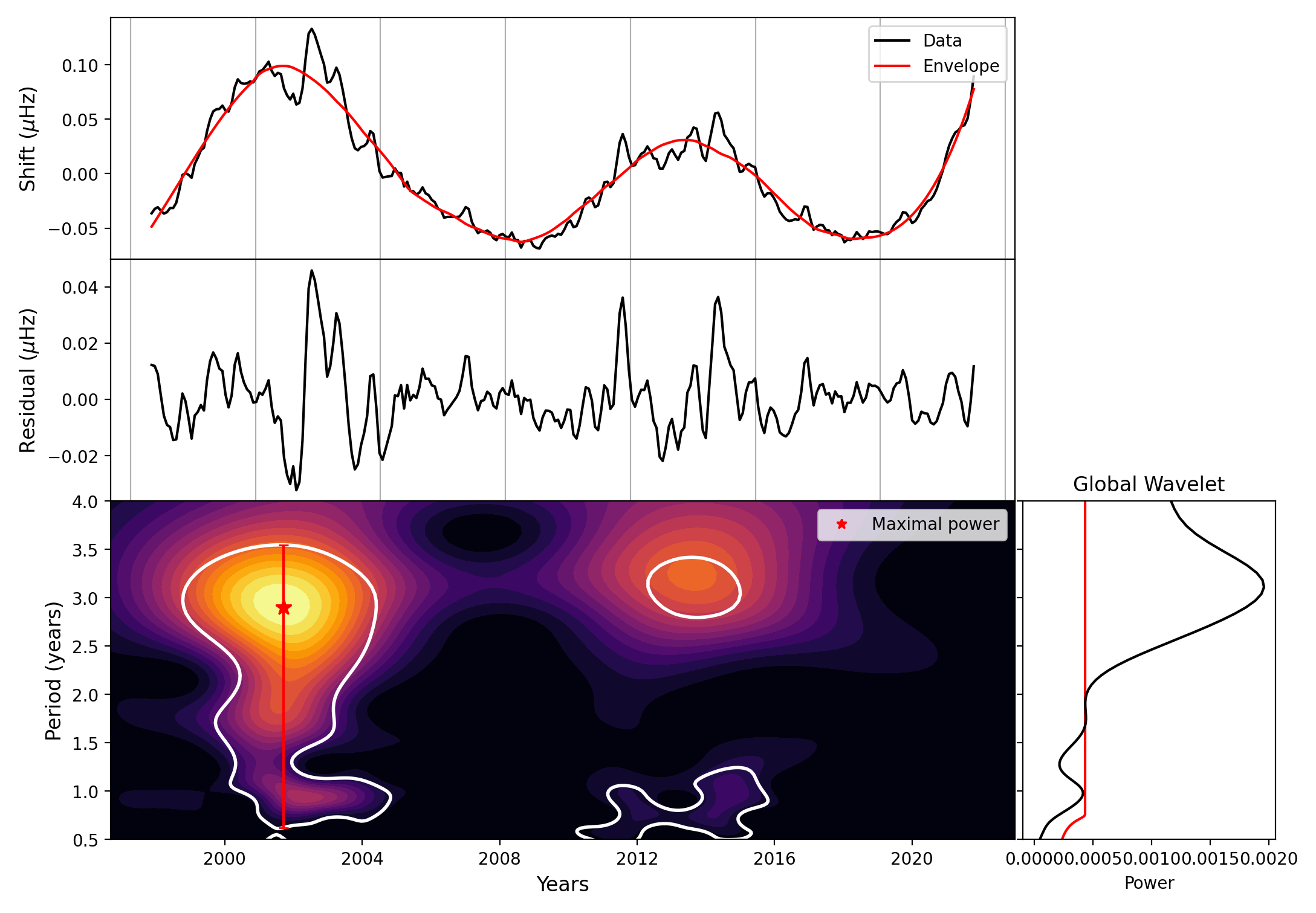}
\caption{Wavelet power spectra for latitude band 80–90}
\label{fig:1900_4100-80_90}
\end{center}
\end{figure}

\bsp	
\label{lastpage}
\end{document}